\documentclass[12pt,a4paper]{article}
\pdfoutput=1
\usepackage{indentfirst}
\usepackage{graphicx}
\usepackage{subfigure}
\usepackage{threeparttable}
\usepackage{array}
\usepackage{multirow}
\usepackage{amsmath}
\usepackage{amssymb}
\usepackage{hyperref}
\usepackage{bm}
\hypersetup{pdfauthor={Jianfeng Chen, Sha Liu, Yong Wang, Chengwen Zhong},pdftitle={A Conserved Discrete Unified Gas-Kinetic Scheme with Unstructured Discrete Velocity Space}}

\begin{document}

%----------------------------------------------------------
% 1. Title
%----------------------------------------------------------

\title{A Conserved Discrete Unified Gas-Kinetic Scheme with Unstructured Discrete Velocity Space}
\renewcommand{\thefootnote}{\fnsymbol{footnote}}
\author{Jianfeng Chen\footnotemark[1], Sha Liu\footnotemark[1], Yong Wang\footnotemark[1], Chengwen Zhong\footnotemark[1]}
\footnotetext[1]{National Key Laboratory of Science and Technology on Aerodynamic Design and Research, Northwestern Polytechnical University, Xi'an, Shaanxi 710072, China}
\footnotetext{\emph{Email addresses:} chenjf@mail.nwpu.edu.cn (Jianfeng Chen), shaliu@nwpu.edu.cn (Sha Liu), wongyung@mail.nwpu.edu.cn (Yong Wang), zhongcw@nwpu.edu.cn (Chengwen Zhong)}
\date{July 13, 2019}
\maketitle

%----------------------------------------------------------
% 2. Abstract
%----------------------------------------------------------

\rule[-5pt]{\textwidth}{0.5pt}
\begin{abstract}
Discrete unified gas-kinetic scheme (DUGKS) is a multi-scale numerical method for flows from continuum limit to free molecular limit, and is especially suitable for the simulation of multi-scale flows, benefiting from its multi-scale property. To reduce integration error of the DUGKS and ensure the conservation property of the collision term in isothermal flow simulations, a Conserved-DUGKS (C-DUGKS) is proposed. On the other hand, both DUGKS and C-DUGKS adopt Cartesian-type discrete velocity space, in which Gaussian and Newton-Cotes numerical quadrature are used for calculating the macroscopic physical variables in low speed and high speed flows, respectively. While, the Cartesian-type discrete velocity space leads to huge computational cost and memory demand. In this paper, the isothermal C-DUGKS is extended to the non-isothermal case by adopting coupled mass and inertial energy distribution functions. Moreover, since the unstructured mesh, such as the triangular mesh in two dimensional case, is more flexible than the structured Cartesian mesh, it is introduced to the discrete velocity space of C-DUGKS, such that more discrete velocity points can be arranged in the velocity regions that enclose large number of molecules, and only a few discrete velocity points need to be arranged in the velocity regions with small amount of molecules in it. By using the unstructured discrete velocity space, the computational efficiency of C-DUGKS is significantly increased. A series of numerical tests in a wide range of Knudsen numbers, such as the Couette flow, lid-driven cavity flow, two-dimensional rarefied Riemann problem and the supersonic cylinder flows, are carried out to examine the validity and efficiency of the present method.
~\\

\end{abstract}
\rule[5pt]{\textwidth}{0.5pt}

%----------------------------------------------------------
% 3. Main text
%----------------------------------------------------------

\section{Introduction}
With the fast development of hypersonic vehicles and near-space aerocraft, numerical flow solvers are required to be applicable to flows in all flow regimes from continuum limit to free molecular limit.  In multi-scale simulations, both numerical methods for continuum flow, such as computational fluid dynamics (CFD) methods based on Navier-Stokes (N-S) equations, and rarefied numerical methods, such as direct simulation Monte Carlo (DSMC) \cite{bird1994molecular}method for Boltzmann equation, should face the problems of model inaccuracy and large computational cost, respectively. Although the CFD methods have achieved great success in continuum flow simulations, it can not be directly extended for rarefied flows with non-equilibrium effect \cite{bird1994molecular,zhu2016discrete}. On the other hand, due to the limitation of cell size and time step, the computational cost of the DSMC method becomes huge in continuum and near-continuum flow simulations \cite{zhu2016discrete}.

In recent years, the numerical schemes based on gas-kinetic theory have achieved a fast development \cite{zhu2016discrete,guo2015discrete,succi2001lattice,xu2001a,xu2013unified,Zhaoli2013Discrete}. As the governing equation of the gas-kinetic theory, theoretically, the Boltzmann equation can be used for flow simulations in all flow regimes \cite{Cercignani2001Rarefied,kremer2010introduction}. The Discrete velocity method (DVM) \cite{Broadwell2006Study,illner1984discrete,Inamuro1990Numerical} is one of the numerical methods for Boltzmann equation and its Bhatnagar-Gross-Krook-type (BGK-type) \cite{Bhatnagar1954A} model equations, in which a discrete velocity space is used. Since most DVM schemes are single-scale methods with split particle transport and collision processes as the DSMC method, their iteration time step should be less than the mean collision time, and their cell size should be less than the mean free path \cite{aristov2012direct,Mieussens2000Discrete}. To address this problem, some asymptotic preserving schemes \cite{filbet2010class,Chen2015A}, such as the Unified Gas-Kinetic Scheme (UGKS) \cite{xu2013unified,xu2010unified,Xu2011An,Xu2012A,Huang2013A} and Discrete Unified Gas-Kinetic Scheme (DUGKS) \cite{zhu2016discrete,guo2015discrete,Zhaoli2013Discrete}, have been proposed. In the algorithm of UGKS, the particle transport and collision processes are coupled by the analytical solution of the BGK-type model equation for updating the discrete distribution functions, as a result, there is no restrictions of the cell size, and the time step can be chosen according to CFL condition. DUGKS has a similar physical process as the UGKS, except the way of calculating the distribution functions at the cell interface. As a results, the numerical flux of distribution functions at the cell interface is obtained from the time integral of the analytical solution of the BGK-type equation in the UGKS algorithm, while, in the algorithm of DUGKS, this numerical flux is obtained from the distribution functions at the half-time step, which can be determined by the numerical characteristic solution of the BGK-type equation. Being the same with the UGKS, the numerical flux of the DUGKS also couples the particle transport and collision processes, while its mathematical form is simple and easy to compute  \cite{zhu2016discrete}. 

Recently, Liu et al. proposed a conserved-DUGKS (C-DUGKS) for isothermal micro-channel flows in all flow regimes \cite{Liu2018A}. Like the original DUGKS, the C-DUGKS also calculates the micro-flux at the cell interface by means of the characteristic-line theory using the auxiliary distribution functions, while the discrete distribution functions and macroscopic conservative variables (mass, momentum and energy) are updated simultaneously to ensure the conservation of  numerical collision term in a time implicitly form which is similar to that of the UGKS method.

Up to now, both the DUGKS and the C-DUGKS use a Cartesian discrete velocity space with either Gauss-Hermite or Newton-Cotes numerical quadrature. For hypersonic rarefied gas flow simulations, a huge number of discrete velocity points should be adopted to capture the non-equilibrium distribution functions accurately, and the corresponding computational cost and memory demand is huge.

To improve the computational efficiency of the asymptotic preserving gas-kinetic methods such as the UGKS and the DUGKS, Zhu et al. and Pan et al. proposed implicit UGKS \cite{Zhu2016Implicit} and implitct DUGKS schemes \cite{pan2017implicit} for non-equilibrium steady flow, respectively, and the multi-grid technology is used to accelerate the convergence \cite{Zhu2017Unified}. Then, the corresponding schemes are extended to unsteady flow simulations further \cite{zhu2019implicit}. Up to now, the implicit scheme with multi-grid technology improves the computational efficiency significantly (up to 100 times faster than the explicit ones). Besides the implicit method with multi-grid technology, the velocity space adaptation  \cite{Chen2012A,arslanbekov2013kinetic,kolobov2014solving,baranger2014locally,brull2016two} is another feasible method for acceleration propose. To improve the computational efficiency and reduce the memory consumption of the UGKS method, Yuan et al. introduced a unstructured discrete velocity space in the numerical scheme \cite{yuan2018conservative}, and the corresponding technology is proposed to suppress the integration error on unstructured mesh. Since the unstructured mesh is flexible, it can be made dense in the regions easily where the value of distribution functions is large, and be made coarse when the value of distribution functions is small. As a result, a small amount of discrete velocity points are sufficient for capturing the non-equilibrium distribution functions.

In this paper, the C-DUGKS method is extended from isothermal and incompressible one to more general non-isothermal and compressible one by adopting the coupled mass and inertial energy distribution functions and the corresponding coupled BGK-type equation system. On the other hand, the unstructured velocity space is introduced to C-DUGKS to improve the computational efficiency, and the validity of the unstructured velocity space is further examined. The Shakhov model is chosen from BGK-type model equations, since its numerical behavior is better than the other models especially for high Mach number flows.

The rest of the paper is organized as follows: the present non-isothermal and compressible C-DUGKS based on the Shakhov model introduced in Section \ref{CDUGKS} along with the unstructured discrete velocity space. In Section \ref{Numerical results}, a series of  numerical tests in different flow regimes, including the Couette flow, lid-driven cavity flow, two-dimensional rarefied Riemann problem and the supersonic cylinder flow,  are used to examine the validity, accuracy and computational efficiency of the present method. The discussion and conclusion  are in Section \ref{Conclusions}.

%%%%%%%%%%%%%%%%%%%%%%%%%%%%%%%%

\section{The conserved discrete unified gas-kinetic scheme}\label{CDUGKS}
\subsection{Shakhov model for Boltzmann equation}\label{Shakhov model}
The starting point of the DUGKS is the BGK-type Boltzmann model equations that use simple relaxation operators instead of complicate Boltzmann collision term. The BGK equation corresponds to a fixed unity Prandtl (Pr) number \cite{guo2015discrete}. To get the right Pr number, several modified models are proposed based on different physical considerations, such as the Shakhov model \cite{Shakhov1968Generalization} and the ellipsoidal statistical model (ES model) \cite{Holway1966New} . The D-dimensional (D is from 1 to 3) Shakhov model equation used in this paper is in the form of 
\begin{equation}\label{equ01}
\frac{{\partial f}}{{\partial t}} + {{\bm\xi }} \cdot \nabla f = \Omega  \equiv  - \frac{{f - f_{}^s}}{\tau },
\end{equation}
where $f = f\left( {{\bm{x}},{{\bm\xi }},{{\bm\eta }},{{\bm\zeta }},t} \right)$ is the velocity distribution functions for particles moving in D-dimensional physical space with velocity ${{\bm\xi }} = \left( {\xi _1^{},...,\xi _D^{}} \right)$  at position ${\bm{x}} = \left( {x_1^{},...,x_D^{}} \right)$ and time $t$. Here ${{\bm\eta }} = \left( {\xi _{D + 1}^{},{\text{ }}.{\text{ }}.{\text{ }}.{\text{ }},\xi _3^{}} \right)$ is the dummy velocity (with the degree of freedom $L = 3 - D$) consisting of the rest components of the translational velocity of particles in three-dimensional space; ${{\bm\zeta }}$ is a vector of $K$ elements representing the internal degree of freedom of molecules; $\Omega $ is the collision operator, $\tau$=$\mu$/$p$, $\mu$ is dynamic viscous coefficient, $p$ is static pressure, and $f_{}^s$ is the Shakhov equilibrium distribution functions in the following form:
\begin{equation}\label{equ02}
f_{}^s = f_{}^{eq}\left[ {1 + \left( {1 - \Pr } \right)\frac{{{\bm{c}} \cdot {\bm{q}}}}{{5pRT}}\left( {\frac{{c_{}^2 + \eta _{}^2}}{{RT}} - 5} \right)} \right] = f_{}^{eq} + f_{\Pr }^{},
\end{equation}
where $f^{eq}$ is the Maxwellian distribution functions, ${\bm{c}} = {{\bm\xi }} - {\bm{u}}$ is the peculiar velocity, ${\bm{u}}$ is the macroscopic flow velocity, ${{\bm{q}}}$ is the heat flux, $R$ is the specific gas constant, and $T$ is the temperature. The Maxwellian distribution functions is given by
\begin{equation}\label{equ03}
f_{}^{eq} = \frac{\rho }{{\left( {2\pi RT} \right)_{}^{\left( {3 + K} \right)/2}}}\exp \left( { - \frac{{{c}{}^2 + \eta _{}^2 + \zeta _{}^2}}{{2RT}}} \right),
\end{equation}
where $\rho $ is the density. In the gas-kinetic theory, the conservative flow variables are defined by the moments of the distribution functions as follows:
\begin{equation}\label{equ04}
{\bm{W}} = \left( {\begin{array}{*{20}{c}}
	\rho  \\
	{\rho {\bm{u}}} \\
	{\rho E}
	\end{array}} \right) = \int {{\bm{\psi }}\left( {{{\bm\xi }},{{\bm\eta }},{{\bm\zeta }}} \right)f} d{{\bm\xi }}d{{\bm\eta }}d{{\bm\zeta }},
\end{equation}
where $\bm\psi  = \left( {{\text{1, }}{{\bm\xi }},\left( {{\xi}^2 + {\eta}^2 + {\zeta}^2} \right)/2} \right)^T$  , $\rho E = \rho u^2/2 + \rho \varepsilon $ is the total energy density, $\rho \varepsilon  = \rho c_V T$  is the inertial energy density, $c_V = \left( {3 + K} \right)R/2$  is the specific heat capacity at constant volume. The specific heat ratio is
\begin{equation}\label{equ05}
\gamma  = \frac{{c_p^{}}}{{c_V^{}}} = \frac{{5 + K}}{{3 + K}},
\end{equation}
where $c_p = \left( {5 + K} \right)R/2$ is the specific heat capacity at constant pressure.
In the gas-kinetic theory, the heat flux is defined by
\begin{equation}\label{equ06}
{\bm{q}} = \frac{1}{2}\int {{\bm{c}}\left( {c_{}^2 + \eta _{}^2 + \zeta _{}^2} \right)f} d{{\bm\xi }}d{{\bm\eta }}d{{\bm\zeta }},
\end{equation}
and the stress tensor is defined by
\begin{equation}\label{equ07}
{{\bm\tau }} = \int {{\bm{cc}}\left( {f - f_{}^{eq}} \right)} d{{\bm\xi }}d{{\bm\eta }}d{{\bm\zeta }}.
\end{equation}

The expression of viscous coefficient $\mu $ is related  to the law of intermolecular interactions \cite{bird1994molecular,guo2015discrete,harris2004introduction}. For variable hard-sphere (VHS) molecules, the viscous coefficient is 
\begin{equation}\label{equ08}
\mu  = \mu _{ref}^{}\left( {\frac{T}{{T_{ref}^{}}}} \right)_{}^\omega ,
\end{equation}
here $\omega $  is the heat index, which is 0.5 for HS model and 0.81 for Argon gas. The relation between the mean free path $\lambda $ and $\mu $ (VHS model) can be written as the follows:
\begin{equation}\label{equ09}
\lambda  = \frac{{2\mu \left( {5 - 2\omega } \right)\left( {7 - 2\omega } \right)}}{{15\rho \left( {2\pi RT} \right)_{}^{1/2}}}.
\end{equation}
In the non-dimensional system, the Knudsen (Kn) number is defined as
\begin{equation}\label{equ10}
Kn = \frac{\lambda }{L} = \frac{{2\mu \left( {5 - 2\omega } \right)\left( {7 - 2\omega } \right)}}{{15\rho \left( {2\pi RT} \right)_{}^{1/2}L}} = \sqrt {\frac{{\gamma }}{\pi }} \frac{{\sqrt {2}\left( {5 - 2\omega } \right)\left( {7 - 2\omega } \right)}}{{15}}\frac{{Ma}}{{\operatorname{Re} }},
\end{equation}
where  $ Ma = \left. {u} / \sqrt {\gamma RT} \right. $ and $\operatorname{Re}  = \left. \rho {u}L / \mu \right. $ are the Mach number and Reynolds number, respectively, and L is the reference length.

%%%%%%%%%%%%%%%%%%%%%%%%%%%%%%%%

\subsection{Reduced distributions}\label{Reduced distribution}
The transport process of molecules depends on the D-dimensional particle velocity ${\bm\xi }$ only, and is irrelevant to ${\bm\eta }$ and ${\bm\zeta }$. To avoid the discretization of ${\bm\eta }$ and ${\bm\zeta }$, two reduced distribution functions \cite{Yang1995Rarefied} are adopted in the present numerical scheme as follows:
\begin{equation}\label{equ11}
\left\{ {\begin{array}{*{20}{c}}
	{g\left( {{\bm{x}},{{\bm\xi }},t} \right) = \int {f\left( {{\bm{x}},{{\bm\xi }},{{\bm\eta }},{{\bm\zeta }},t} \right)d{{\bm\eta }}d{{\bm\zeta }}} }, \\
	{h\left( {\bm{x},\bm\xi ,t} \right) = \int {\left( {{{\eta }}_{}^2 + {{\zeta }}_{}^2} \right)f\left( {{\bm{x}},{{\bm\xi }},{{\bm\eta }},{{\bm\zeta }},t} \right)d{{\bm\eta }}d{{\bm\zeta }}} .}
	\end{array}} \right.
\end{equation}
Given Eq.(\ref{equ01}) and Eq.(\ref{equ11}), the following BGK-type equation system can be obtained,
\begin{equation}\label{equ12}
\left\{ {\begin{array}{*{20}{c}}
	{\frac{{\partial g}}{{\partial t}} + {\bm{\xi }} \cdot \nabla g = \Omega _g^{} \equiv  - \frac{{g - g_{}^s}}{\tau }}, \\
	{\frac{{\partial h}}{{\partial t}} + {\bm{\xi }} \cdot \nabla h = \Omega _h^{} \equiv  - \frac{{h - h_{}^s}}{\tau }},
	\end{array}} \right.
\end{equation}
where the reduced Shakhov equilibrium distribution functions $g_{{}}^{s}$ and $h_{{}}^{s}$ are given by
\begin{equation}\label{equ13}
\left\{ {\begin{array}{*{20}{c}}
	{g_{}^s = \int {f_{}^s\left( {{\bm{x}},{{\bm\xi }},{{\bm\eta }},{{\bm\zeta }},t} \right)d{{\bm\eta }}d{{\bm\zeta }}}  = g_{}^{eq} + g_{\Pr }^{}}, \\
	{h_{}^s = \int {\left( {\eta _{}^2 + \zeta _{}^2} \right)f_{}^s\left( {{\bm{x}},{{\bm\xi }},{{\bm\eta }},{{\bm\zeta }},t} \right)d{{\bm\eta }}d{{\bm\zeta }}}  = h_{}^{eq} + h_{\Pr }^{}},
	\end{array}} \right.
\end{equation}
with
\begin{equation}\label{equ14}
g_{}^{eq} = \int {f_{}^{eq}d{{\bm\eta }}d{{\bm\zeta }}}  = \frac{\rho }{{\left( {2\pi RT} \right)_{}^{D/2}}}\exp \left( { - \frac{{c_{}^2}}{{2RT}}} \right),
\end{equation}
\begin{equation}\label{equ15}
g_{\Pr }^{} = \int {f_{\Pr }^{}d{{\bm\eta }}d{{\bm\zeta }}}  = \left( {1 - \Pr } \right)\frac{{{\bm{c}} \cdot {\bm{q}}}}{{5pRT}}\left( {\frac{{c_{}^2}}{{RT}} - D - 2} \right)g_{}^{eq},
\end{equation}
\begin{equation}\label{equ16}
h_{}^{eq} = \int {\left( {\eta _{}^2 + \zeta _{}^2} \right)f_{}^{eq}d{{\bm\eta }}d{{\bm\zeta }}}  = \left( {K + 3 - D} \right)RTg_{}^{eq},
\end{equation}
\begin{equation}\label{equ17}
h_{\Pr }^{} = \int {\left( {\eta _{}^2 + \zeta _{}^2} \right)f_{\Pr }^{}d{{\bm\eta }}d{{\bm\zeta }}}  = \left( {1 - \Pr } \right)\frac{{{\bm{c}} \cdot {\bm{q}}}}{{5pRT}}\left[ {\left( {\frac{{c_{}^2}}{{RT}} - D} \right)\left( {K + 3 - D} \right) - 2K} \right]RTg_{}^{eq}.
\end{equation}
\par By substituting Eq.(\ref{equ11}) into Eq.(\ref{equ04}), (\ref{equ06}), (\ref{equ07}), it can be obtained that
\begin{equation}\label{equ18}
\left\{ {\begin{array}{*{20}{c}}
	{\rho  = \int {gd{{\bm\xi }}} }, \\
	{\rho {\bm{u}} = \int {{{\bm\xi }}gd{{\bm\xi }}} }, \\
	{\rho E = \frac{1}{2}\int {\left( {{{\xi }}_{}^2g + h} \right)d{{\bm\xi }}} },
	\end{array}} \right.
\end{equation}
\begin{equation}\label{equ19}
{\bm{q}} = \frac{1}{2}\int {{\bm{c}}\left( {c_{}^2g + h} \right)} d{{\bm\xi }},
\end{equation}
\begin{equation}\label{equ20}
{{\bm\tau }} = \int {{\bm{cc}}\left( {g - g_{}^{eq}} \right)} d{{\bm\xi }}.
\end{equation}
\par Using the reduced distribution functions, the conservation property of Shakhov collision operator can also be written as:
\begin{equation}\label{equ21}
\left\{ {\begin{array}{*{20}{c}}
	{\int {\Omega _g^{}d{{\bm\xi }} = 0} }, \\
	{\int {{{\bm\xi }}\Omega _g^{}d{{\bm\xi }} = 0} }, \\
	{\int {\left( {{{\xi }}_{}^2\Omega _g^{} + \Omega _h^{}} \right)d{{\bm\xi }} = 0} }.
	\end{array}} \right.
\end{equation}

%%%%%%%%%%%%%%%%%%%%%%%%%%%%%%%%

\subsection{The Conserved DUGKS}\label{aCDUGKS}
As can be seen from Eq.(\ref{equ12}), $g$ and $h$  have the same updating rule, therefor they can be replaced by a new symbol $\phi $ in the algorithm for simplicity, and the governing equation can be written as:
\begin{equation}\label{equ22}
\frac{{\partial \phi }}{{\partial t}} + {{\bm\xi }} \cdot \nabla \phi  = \Omega _\phi ^{} \equiv  - \frac{{\phi  - \phi _{}^s}}{\tau }.
\end{equation}

Integrating Eq.(\ref{equ22}) in control volume $V_j^{}$ from time $t_n$ to $t_{n + 1}{\text{ = }}t_n{\text{ + }}\Delta t$, it can obtain that
\begin{equation}\label{equ23}
\phi _j^{n + 1}\left( {\bm{\xi }} \right) - \phi _j^n\left( {\bm{\xi }} \right) + \frac{{\Delta t}}{{\left| {V_j^{}} \right|}}F_j^{n + 1/2}\left( {\bm{\xi }} \right) = \frac{{\Delta t}}{2}\left[ {\Omega _j^{n + 1}\left( {\bm{\xi }} \right) + \Omega _j^n\left( {\bm{\xi }} \right)} \right],
\end{equation}
where $\left| V_j \right|$ is the volume of $V_j$  and $F_{}^{n + 1/2}\left( {{\bm\xi }} \right)$ is the microflux across the cell interface given by
\begin{equation}\label{equ24}
F_j^{n + 1/2}\left( {{\bm\xi }} \right) = \int_{\partial V_j^{}} {\left( {{\bm{\xi }} \cdot {\bm{n}}} \right)} \phi \left( {\bm{x},{\bm{\xi }},t_{n + 1/2}^{}} \right)d\bm{S},
\end{equation}
where $\partial V_j$ is the surface of the control volume  $V_j^{}$ and  ${\bm{n}}$ is the external normal unit vector of $d\bm{S}$ (differential of $V_j^{}$). It should be noted that a trapezoidal rule is used for the time discretization of the collision term.

In order to avoid calculating the collision term at $t_{n + 1}$  time on the right side of Eq.(\ref{equ23}), DUGKS introduces two new auxiliary distribution functions. While, the conservative variables of C-DUGKS are updated first \cite{Liu2018A}, so that the  equilibrium distribution functions at $t_{n + 1}$ can be obtained, then the implicit evolution equation can be transformed into the following explicit one,
\begin{equation}\label{equ25}
\begin{aligned}
\phi _j^{n + 1}\left( {\bm{\xi }} \right) = \left( {1 + \frac{{\Delta t}}{{2\tau _j^{n + 1}}}} \right)_{}^{ - 1}\left[ {\phi _j^n\left( {\bm{\xi }} \right) - \frac{{\Delta t}}{{\left| {V_j^{}} \right|}}F_j^{n + 1/2}\left( {\bm{\xi }} \right) +} \right. \\ 
\left. {\frac{{\Delta t}}{2}\left( {\frac{{\phi _j^{s,n + 1}\left( {\bm{\xi }} \right)}}{{\tau _j^{n + 1}}} + \frac{{\phi _j^{s,n}\left( {\bm{\xi }} \right) - \phi _j^n\left( {\bm{\xi }} \right)}}{{\tau _j^n}}} \right)} \right].
\end{aligned}
\end{equation}

Similar to the UGKS, conservative variables ${\bm{W}_j}^{n + 1}$ need to be updated before calculating the equilibrium distribution functions $\phi _j^{s,n + 1}\left( {\bm{\xi }} \right)$ and the relaxation time $\tau _j^{n + 1}$ in Eq.(\ref{equ25}). Given Eq.(\ref{equ18}), Eq.(\ref{equ21}) and Eq.(\ref{equ23}), ${\bm{W}_j}^{n + 1}$ can be updated using the following equations:
\begin{equation}\label{equ26}
\begin{aligned}
\left( {\begin{array}{*{20}{c}}
	{\rho _j^{n + 1}} \\
	{\left( {\rho {\bm{u}}} \right)_j^{n + 1}} \\
	{\left( {\rho E} \right)_j^{n + 1}}
	\end{array}} \right)\\ =  \left( {\begin{array}{*{20}{c}}
	{\rho _j^n} \\
	{\left( {\rho {\bm{u}}} \right)_j^n} \\
	{\left( {\rho E} \right)_j^n}
	\end{array}} \right) -
& \frac{{\Delta t}}{{\left| {V_j^{}} \right|}}\int {} \int_{\partial V_j^{}} {\left( {{{\bm\xi }} \cdot {\bm{n}}} \right)} \left( {\begin{array}{*{20}{c}}
	{g\left( {\bm{x},{{\bm\xi }},t_{n + 1/2}^{}} \right)} \\
	{{{\bm\xi }}g\left( {\bm{x},{{\bm\xi }},t_{n + 1/2}^{}} \right)} \\
	{\xi _{}^2g\left( {\bm{x},{{\bm\xi }},t_{n + 1/2}^{}} \right) + h\left( {\bm{x},{{\bm\xi }},t_{n + 1/2}^{}} \right)}
	\end{array}} \right)dSd{{\bm\xi }.}
\end{aligned}
\end{equation}

Eq.(\ref{equ25}) and Eq.(\ref{equ26}) are the evolution equations for the microscopic distribution functions and the macroscopic conservative variables, respectively. Once the distribution functions $\phi\left(\bm{x},{\bm\xi },t_{n+1/2}^{{}} \right)$ at the half-time step and at the cell interface is obtained, the whole scheme can be  established.

\begin{figure}
	\centering
	\includegraphics[width=0.6\textwidth]{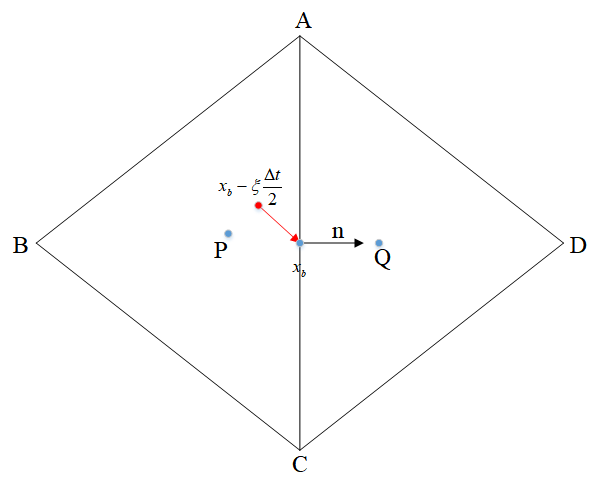}
	\caption{\label{fig:fig01}Sketch of two neighboring cells and the characteristic-line.}
\end{figure}

In order to obtain $\phi\left(\bm{x},\bm{\xi },t_{n+1/2}^{{}} \right)$ , the Shakhov model equation Eq.(\ref{equ22}) is integrated along the characteristic-line (in the direction of particle velocity) from $t_{n}^{{}}$ to $t_{n+1/2}^{{}}$. As shown in Fig. \ref{fig:fig01}, the characteristic  ends at the midpoint of the cell interface. This process can be expressed by the following equation:
\begin{equation}\label{equ27}
\phi^{}\left( {{\bm{x}}_b^{},{\bm{\xi }},t_n^{} + h} \right) - \phi^{}\left( {{\bm{x}}_b^{} - {\bm{\xi }}h,{\bm{\xi }},t_n^{}} \right) = \frac{h}{2}\left[ {\Omega^{}\left( {{\bm{x}}_b^{},{\bm{\xi }},t_n^{} + h} \right) + \Omega^{}\left( {{\bm{x}}_b^{} - {\bm{\xi }}h,{\bm{\xi }},t_n^{}} \right)} \right],
\end{equation}
where ${\bm{x}}_b$ is the midpoint of the cell interface, and $h = \Delta t/2$ is the half-time step. By introducing the following auxiliary distribution functions:
\begin{equation}\label{equ28}
\begin{gathered}
\mathop {\phi _{}^{}}\limits^ -   = \phi  - \frac{h}{2}\Omega  = \frac{{2\tau  + h}}{{2\tau }}\phi  - \frac{h}{{2\tau }}\phi _{}^s \hfill ,\\
\;\mathop {\phi _{}^ + }\limits^ -   = \phi  + \frac{h}{2}\Omega  = \frac{{2\tau  - h}}{{2\tau }}\phi  + \frac{h}{{2\tau }}\phi _{}^s \hfill ,\\
\end{gathered}
\end{equation}
Eq.(\ref{equ27}) can be written as
\begin{equation}\label{equ29}
\mathop {\phi^{}}\limits^ -  \left( {{\bm{x}}_b^{},{\bm{\xi }},t_n^{} + h} \right) = \mathop {\phi^ + }\limits^ -  \left( {{\bm{x}}_b^{} - {\bm{\xi }}h,{\bm{\xi }},t_n^{}} \right).
\end{equation}
That means as long as $\mathop {\phi^ + }\limits^ -  \left( {{\bm{x}}_b - {\bm{\xi }}h,{\bm{\xi }},t_n^{}} \right)$ is known, $\mathop {\phi}\limits^ -  \left( {{\bm{x}}_b,{\bm{\xi }},t_n + h} \right)$ can be directly abtained. Since the collision operator fulfills the conservation property (Eq.(\ref{equ21})), the macroscopic variables at the cell interface at half-time step can be obtained from the  auxiliary distribution functions as follows:
\begin{equation}\label{equ30}
\left\{ {\begin{array}{*{20}{c}}
	{\rho  = \int {\mathop g\limits^ -  d{\bm{\xi }}} }, \\
	{\rho {\bm{u}} = \int {{\bm{\xi }}\mathop g\limits^ -  d{\bm{\xi }}} }, \\
	{\rho E = \frac{1}{2}\int {\left( {{{\xi }}_{}^2\mathop g\limits^ -   + \mathop h\limits^ -  } \right)d{\bm{\xi }}} },
	\end{array}} \right.
\end{equation}
\begin{equation}\label{equ31}
\mathop {\bm{q}}\limits^ -   = \frac{1}{2}\int {{\bm{c}}\left( {c_{}^2\mathop g\limits^ -   + \mathop h\limits^ -  } \right)} d{\bm{\xi }},\;\;{\bm{q}} = \frac{{2\tau }}{{2\tau  + h\Pr }}\mathop {\bm{q}}\limits^ -  \;.
\end{equation}At the half-time step, the Shakhov equilibrium distribution functions $\phi^s\left( {{\bm{x}}_b,{\bm{\xi }},t_n + h} \right)$ at the cell interface can be obtained by calculating the macroscopic conservative variables ${\bm{W}}^{}\left( {{\bm{x}}_b^{},{\bm{\xi }},t_n^{} + h} \right)$  and the heat flux ${\bm{q}}^{}\left( {{\bm{x}}_b^{},{\bm{\xi }},t_n^{} + h} \right)$ at the cell interface. Thus, the distribution functions at the cell interface can be recovered through the following equation:
\begin{equation}\label{equ32}
\phi \left( {{\bm{x}}_b^{},{\bm{\xi }},t_n^{} + h} \right) = \frac{{2\tau }}{{2\tau  + h}}\mathop {\phi _{}^{}}\limits^ -  \left( {{\bm{x}}_b^{},{\bm{\xi }},t_n^{} + h} \right) + \frac{h}{{2\tau  + h}}\phi _{}^s\left( {{\bm{x}}_b^{},{\bm{\xi }},t_n^{} + h} \right).
\end{equation}
\par Since both distribution functions and macroscopic variables at the cell interface at half-step are constructed from  $\mathop {\phi}\limits^ -  \left( {{\bm{x}}_b,{\bm{\xi }},t_n + h} \right)$, and  $\mathop {\phi}\limits^ -  \left( {{\bm{x}}_b,{\bm{\xi }},t_n + h} \right)$ is equal to $\mathop {\phi^ + }\limits^ -  \left( {{\bm{x}}_b - {\bm{\xi }}h,{\bm{\xi }},t_n^{}} \right)$ (Eq.(\ref{equ29})). Therefore, the calculation of  $\mathop {\phi^ + }\limits^ -  \left( {{\bm{x}}_b - {\bm{\xi }}h,{\bm{\xi }},t_n^{}} \right)$ is very important in the present scheme. By using the Taylor expansion at the center of the control volume,  $\mathop {\phi^ + }\limits^ -  \left( {{\bm{x}}_b - {\bm{\xi }}h,{\bm{\xi }},t_n^{}} \right)$ can be obtained from the following reconstruction:
\begin{equation}\label{equ33}
\begin{aligned}
\mathop {\phi^ + }\limits^ -  \left( {{\bm{x}}_b^{} - {\bm{\xi }}h,{\bm{\xi }},t_n^{}} \right) = \mathop {\phi^ + }\limits^ -  \left( {{\bm{x}}_c^{},{\bm{\xi }},t_n^{}} \right) + \\ \Psi \left( {{\bm{x}}_c^{},{\bm{\xi }},t_n^{}} \right)\nabla \mathop {\phi^ + }\limits^ -  \left( {{\bm{x}}_c^{},{\bm{\xi }},t_n^{}} \right) \cdot \left( {{\bm{x}}_b^{} - {\bm{\xi }}h - {\bm{x}}_c^{}} \right),  
& \;{\bm{x}}_b^{} - {\bm{\xi }}h \in V_c^{},
\end{aligned}
\end{equation}
where $V_c$ stands for control volume which is centered at point $C$ (Fig. \ref{fig:fig01}). If ${\bm{\xi }} \cdot {\bm{n}}_b^{} \geqslant 0$, point $C$ is  $P$ (the center the left cell) in Fig. \ref{fig:fig01}, otherwise point $C$ is $Q$ (the center of the right cell). $\nabla \mathop {\phi^ + } \left( {{\bm{x}}_c^{},{\bm{\xi }},t_n^{}} \right)$ is the gradient of the auxiliary distribution functions at point $C$, which is calculated by the least square method in this paper, and $\Psi \left( {{\bm{x}}_c^{},{\bm{\xi }},t_n^{}} \right)$ is the gradient limiter used to suppress the numerical oscillations. The Venkatakrishnan  limiter \cite{Venkatakrishnan1995Convergence} is  chosen in this paper.

The time step of the present explicit scheme is determined by the following Courant-Friedrichs-Lewy (CFL) condition,
\begin{equation}\label{equ34}
\Delta t = \alpha \frac{{\Delta x}}{{{\bm{\xi }} + {\bm{u}}}},
\end{equation}
where $\alpha $ is the CFL number, and $\Delta x$ is the minimum grid spacing.

\begin{figure}
	\centering
	\includegraphics[width=0.6\textwidth]{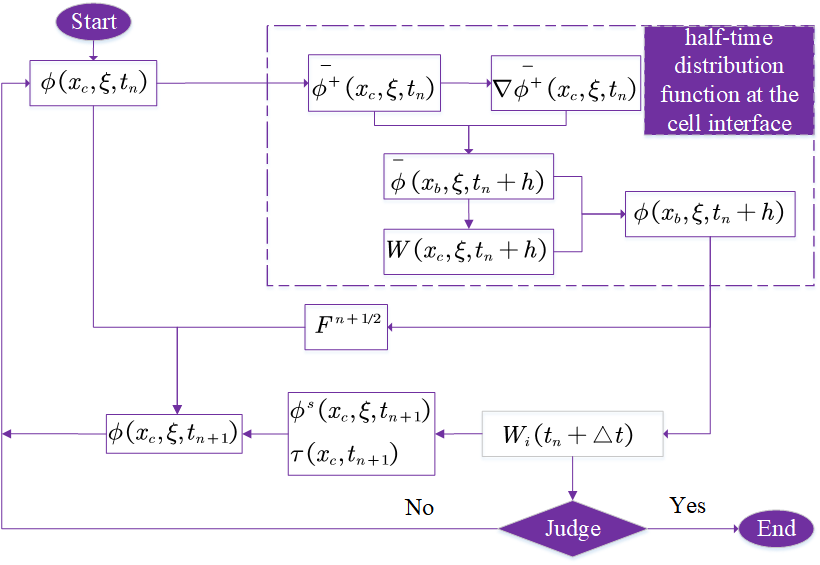}
	\caption{\label{fig:fig02}Flow chart of C-DUGKS.}
\end{figure}

Fig. \ref{fig:fig02} shows the flow chart of the present compressible and non-isothermal C-DUGKS. The key point of the present scheme is calculating the distribution functions at the cell interface at the half-time step.

%%%%%%%%%%%%%%%%%%%%%%%%%%%%%%%%

\subsection{Unstructured discrete velocity space}\label{UDVS}
By integrating the distribution functions in the continuous velocity space, the conservative variables can be obtained from Eq.(\ref{equ18}). In numerical simulations, when discrete velocity space is adopted, the continuous integration in Eq.(\ref{equ18}) is replaced by the following numerical quadrature:
\begin{equation}\label{equ35}
\left\{ {\begin{array}{*{20}{c}}
	{\rho  = \sum\limits_{i = 1}^b {w_i^{}g\left( {{\bm{\xi }}_i^{}} \right)} }, \\
	{\rho {\bm{u}} = \sum\limits_{i = 1}^b {w_i^{}{\bm{\xi }}_i^{}g\left( {{\bm{\xi }}_i^{}} \right)} }, \\
	{\rho E = \frac{1}{2}\sum\limits_{i = 1}^b {w_i^{}\left( {{\mathbf{\xi }}_i^2g\left( {{\bm{\xi }}_i^{}} \right) + h\left( {{\bm{\xi }}_i^{}} \right)} \right)}. }
	\end{array}} \right.
\end{equation}
Up to now, the DUKGS generally uses Gauss-Hermite  numerical quadrature for low-speed continuum flows and Newton-Cotes  numerical quadrature for high-speed rarefied flows. These two  numerical quadrature are conducted on structured Cartesian mesh in velocity space. Therefore, when the flow Mach number is large, the amount of discrete velocity points need to be extremely large. Actually, according to Eq.(\ref{equ35}), the velocity mesh should be dense only in the regions where the value of distribution function is large, and can be coarse in the regions with small distribution function.

\begin{figure}
	\centering
	\subfigure[Structured velocity mesh]{\includegraphics[width=0.4\textwidth]{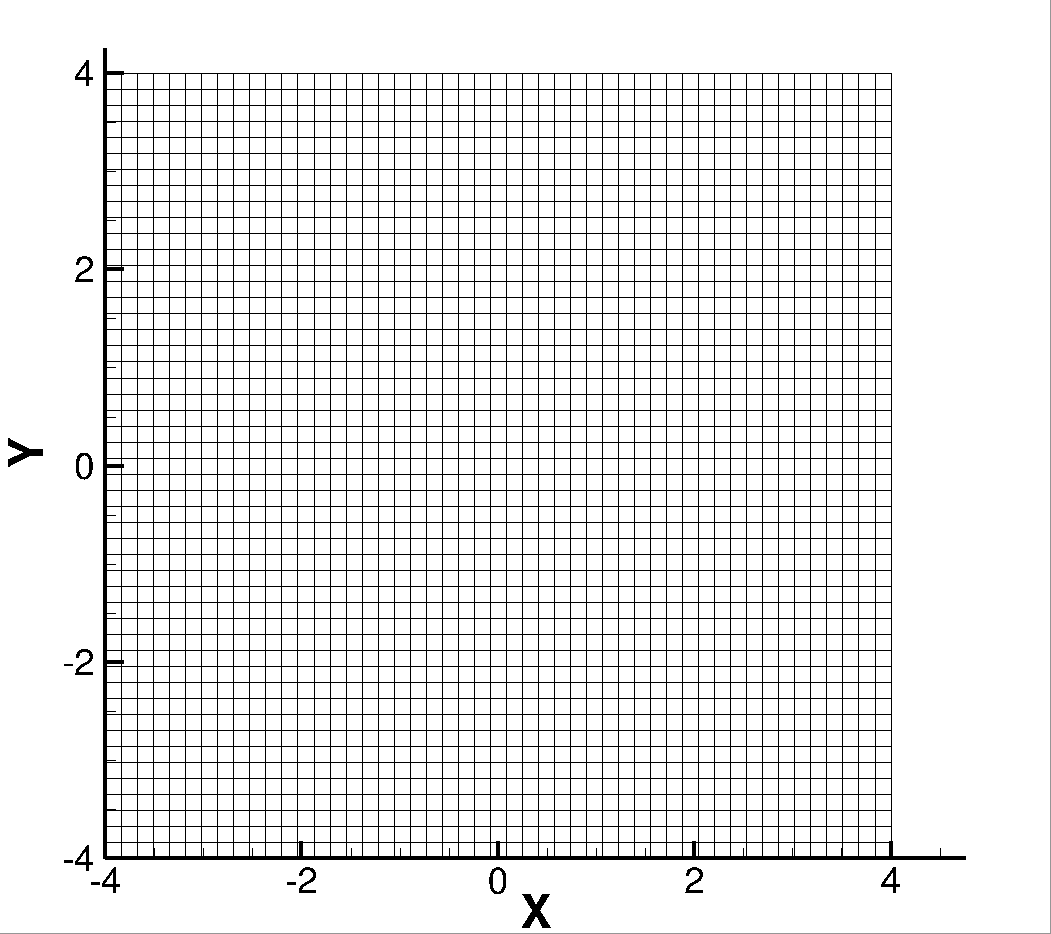}}
	\subfigure[Unstructured velocity mesh]{\includegraphics[width=0.4\textwidth]{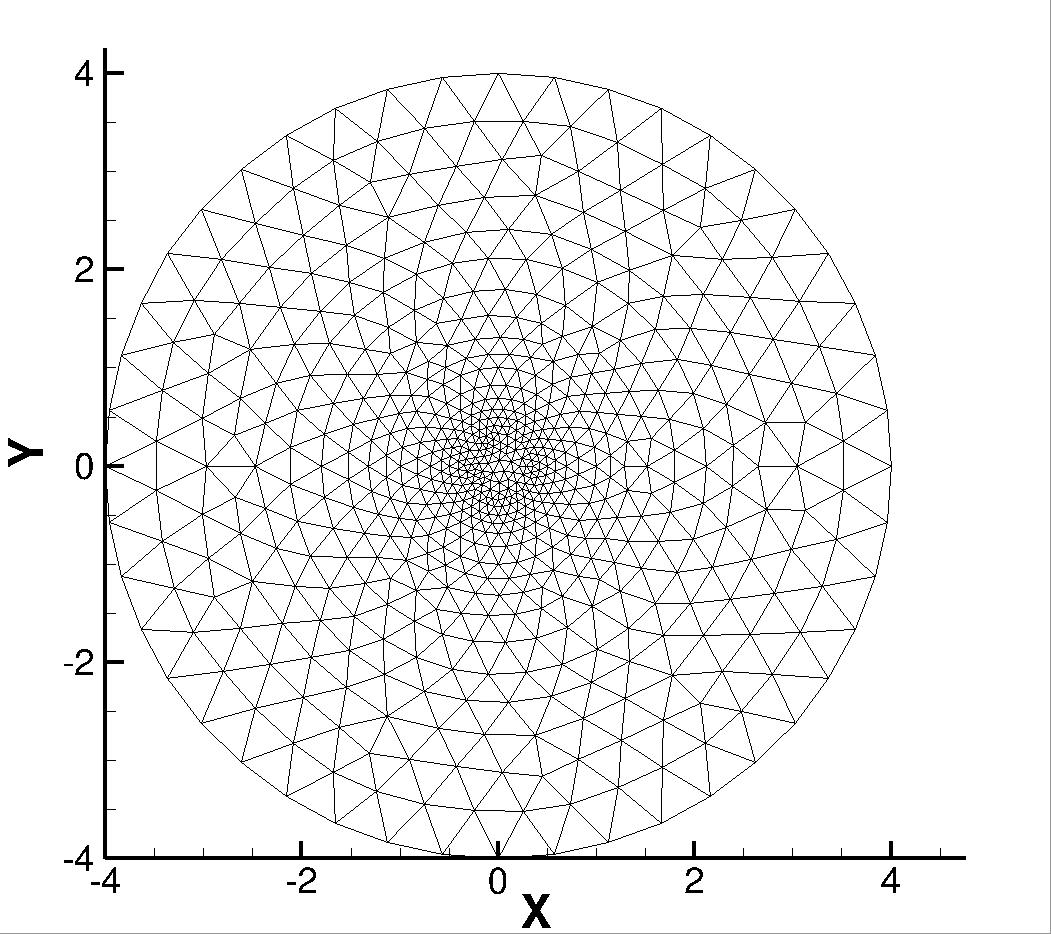}}
	\caption{\label{fig:fig03}Discrete velocity spaces.}
\end{figure}

To reduce the amount of discrete velocity points and improve the computational efficiency, Yuan et al.  introduced the unstructured discrete velocity into the UGKS method, since the unstructured mesh (the triangular mesh for example) is more flexible than the structured one \cite{yuan2018conservative}. Fig. \ref{fig:fig03} illustrates a simple comparison between the structured Cartesian mesh and unstructured triangular mesh for discrete velocity space about a Maxwellian distribution with zero macroscopic speed.  The  triangular mesh can be dense at the zero velocity point where the distribution functions is the maximum, and gradually become coarse when the molecular speed is large (far from the zero velocity point in velocity space). Obviously, the unstructured discrete velocity space  is flexible. It can refine and coarsen the mesh when needed. This makes the unstructured mesh a efficient way in organizing the discrete velocity space.

In the unstructured discrete velocity space, the discrete points and their weights are the central point and the area of mesh cells, respectively. Theoretically, the accuracy of this type of integration is slightly lower than that of Gauss-Hermite and Newton-Cotes ones on structured Cartesian mesh. While, since the unstructured mesh is flexible, its integration accuracy is often higher than that of the structured one with the same amount of discrete points. In this paper, the unstructured discrete velocity space is introduced to the C-DUGKS for the first time to improve the computational efficiency and reduce memory cost. Since DUGKS has a similar physical process as UGKS, it is easy to believe that unstructured discrete velocity space will work well in C-DUGKS. Several examples will be carried out to explore the performance of present method in the next section.

%%%%%%%%%%%%%%%%%%%%%%%%%%%%%%%%	

\section{Numerical experiments}\label{Numerical results}
A series of test cases are carried out using the present C-DUGKS with both structured Cartesian discrete velocity space and unstructured triangular discrete velocity space. The accuracy and numerical efficiency of the two methods are compared. The results of UGKS and DSMC are chosen as the benchmark solutions. 

The first test case is the Couette flow with Kn number $0.01$, ${\text{0}}{\text{.2/}}\sqrt \pi  $, ${\text{2/}}\sqrt \pi  $ and ${\text{20/}}\sqrt \pi  $. The second test case is the lid-driven cavity flow with Kn number $0.075$, $1$ and $10$. The third test case is the two-dimensional rarefied Riemann problem.  The last test case is the supersonic cylinder flow with Kn number $0.1$ and $1$. 
For simulation of steady flow in this paper, the relative errors of macroscopic variables are calculated every 1000 steps. When the relative error is less than $\varepsilon $ as the following,
\begin{equation}\label{equ36}
\frac{{\sqrt {\sum {{{\left( {{{\bm{u}}^{n + 1000}}{\text{ - }}{{\bm{u}}^n}} \right)}^{\text{2}}}} } }}{{\sqrt {\sum {{{\left( {{{\bm{u}}^{n + 1000}}} \right)}^{\text{2}}}} } }} < \varepsilon ,
\end{equation}
it is considered that the flow has reached the stable state. In this paper, $\varepsilon $ is chosen as ${10^{ - 6}}$ for all steady test cases. All numerical simulations are conducted on a compute node with 24 cores, and the CPU information is Intel(R) Xeon(R) CPU E5-2685 v3 @ 2.6GHz.

\subsection{Couette flow}\label{Couette flow}
\begin{figure}
	\centering
	\includegraphics[width=0.6\textwidth]{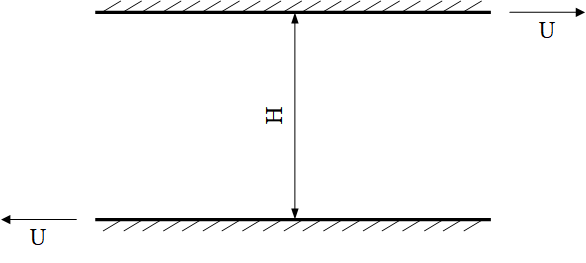}
	\caption{\label{fig:fig04}Geometry of the Couette flow.}
\end{figure}

The Couette flow is driven by two parallel plates whose distance is $H$ as shown in Fig. \ref{fig:fig04}. The upper and lower plates move in the opposite directions with a magnitude $U_w$. In this case, the physical space between the two plates is discretized into 100 cells. The working gas is Argon ($R=208J/kg/K$), and HS model is adopted ($\omega  = 0.5$). Taking $H$ as the reference length, four cases with Kn number $0.01$, ${\text{0}}{\text{.2/}}\sqrt \pi  $, ${\text{2/}}\sqrt \pi  $ and ${\text{20/}}\sqrt \pi  $ are carried out. The flow field is in stationary initially. Both the initial temperature  and wall temperatures are $273K$. The velocity of plates are  $U_{w}^{{}}=0.5\sqrt{RT}=119.15m/s$. For simplicity, dimensionless physical variables are used. The reference temperature and velocity are $T_{ref}^{{}}=273K$ and $U_{ref}^{{}}=\sqrt{2RT_{ref}^{{}}}=337.00m/s$, respectively. The dimensionless density and temperature of the initial flow are $\rho _{0}^{{}}=1.0$ and $T_{0}^{{}}=1.0$ respectively, and the magnitude of the dimensionless velocity of both plates are $U=0.3535$.

\begin{figure}
	\centering
	\includegraphics[width=0.6\textwidth]{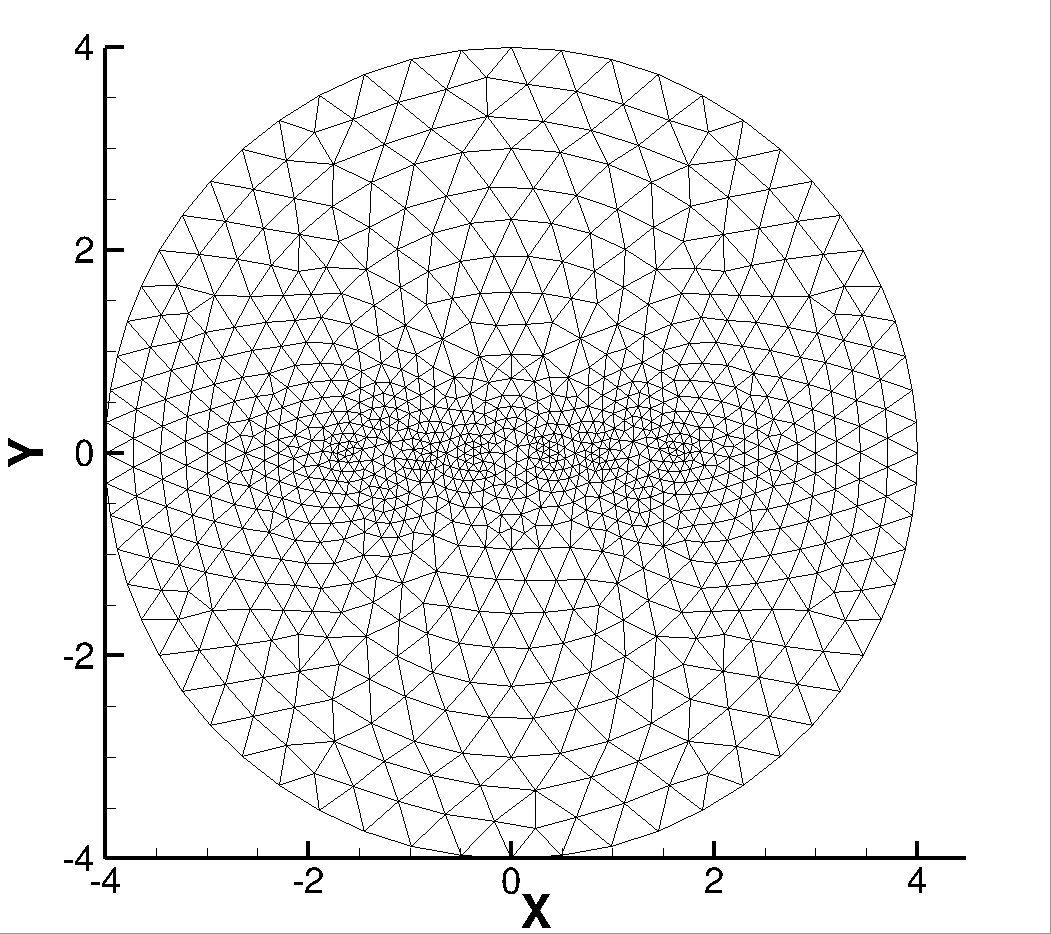}
	\caption{\label{fig:fig07} Unstructured velocity mesh  for Couette flow with 1882 points.}
\end{figure}

Similar to Ref. \cite{Zhu2016Implicit}, $80 \times 80$ uniform discrete velocity points in a range of  $\left[ { - 4,\;4} \right] \times \left[ { - 4,\;4} \right]$ are chosen as the structured discrete velocity space. And the unstructured discrete velocity space shown in Fig. \ref{fig:fig07} is adopted. The range of this unstructured velocity space is a  circle surface that is centered at $\left( {0,0} \right)$ with a radius of 4. The number of discrete velocity points is $1,681$. Although different types of discrete velocity spaces are chosen, they can use the same CFL number, since their minimum spatial mesh size and maximum discrete velocity are the same.

\begin{figure}
	\centering
	\includegraphics[width=0.6\textwidth]{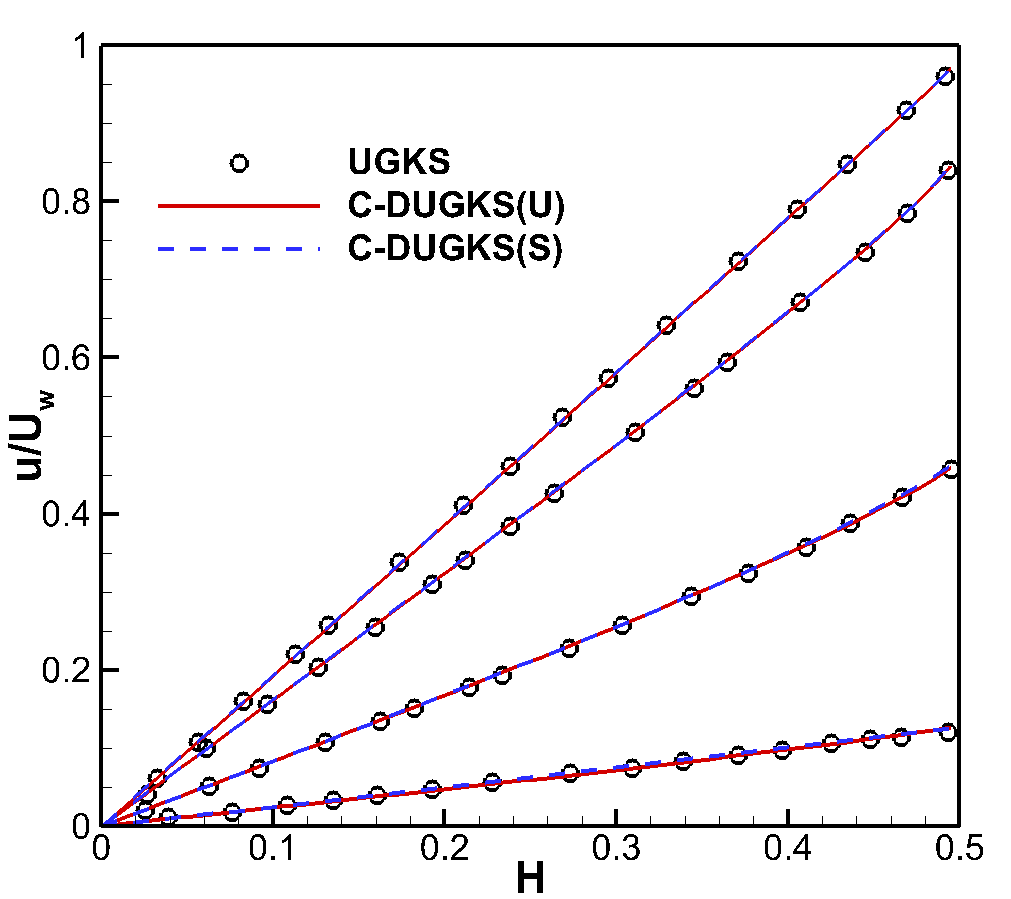}
	\caption{\label{fig:fig08} U-Velocity profile in veritical direction predicted by UGKS and C-DUGKS with unstructured velocity mesh (C-DUGKS(U)) and structured velocity mesh (C-DUGKS(S)).}
\end{figure}

Fig. \ref{fig:fig08} shows the horizontal velocity profile in the vertical direction predicted by C-DUGKS with both structured and unstructured velocity spaces, along with the benchmark UGKS data. Since the vertical velocity is almost zero  for Couette flow, the  discrete velocity spaces in the vertical direction is set coarse, and it is refined in the horizontal direction. It is found that the results predicted by the UGKS and the present C-DUGKS are quite consistent with each other. Since the unstructured mesh is flexible, 1882 discrete velocity points are sufficient in this case, while the number of discrete velocity points are 6400 (80*80) for the structured velocity space. Table \ref{tab:Couette} is the convergence time of C-DUGKS with both structured and unstructured velocity space. Since the computation time is proportional to the number of discrete velocity points, the utilization of the unstructured velocity space leads to a 3 to 4 times speedup.

\begin{table}\label{tab:Couette}	
	\centering
	\caption{Convergence of C-DUGKS with both structured and unstructured velocity meshes (in seconds) for Courtte flow.}
	\begin{tabular}{p{142pt} p{60pt}<{\raggedleft} p{60pt}<{\raggedleft} p{30pt}<{\raggedleft}  }
		\hline
		\hline
		& Structured mesh (A) & Unstructured mesh (B) & ratio (A/B)\\
		\hline
		Discrete velocity points & 6400 & 1882 & 3.4\\
		\hline
		Convergence time (Kn=0.01) &  137.0 & 40.3 & 3.4 \\
		\hline
		Convergence time (Kn=${\text{0.2/}}\sqrt \pi    $) &  36.3	 & 9.3 & 3.9 \\
		\hline
		Convergence time ((Kn=${\text{2/}}\sqrt \pi    $) &   28.1   & 9.3 & 3.0 \\
		\hline
		Convergence time ((Kn=${\text{20/}}\sqrt \pi    $) &  271.7  & 96.1 & 2.8 \\
		\hline
		\hline
	\end{tabular}
\end{table}

\subsection{Cavity flow}\label{cavity flow}

\begin{figure}
	\centering
	\includegraphics[width=0.6\textwidth]{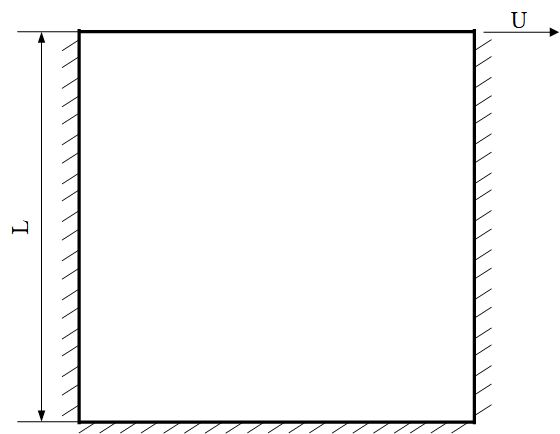}
	\caption{\label{fig:fig09}Geometry of the cavity flow.}
\end{figure}

\begin{figure}
	\centering
	\subfigure[Spatial mesh]{\label{fig:fig10a}\includegraphics[width=0.47\textwidth]{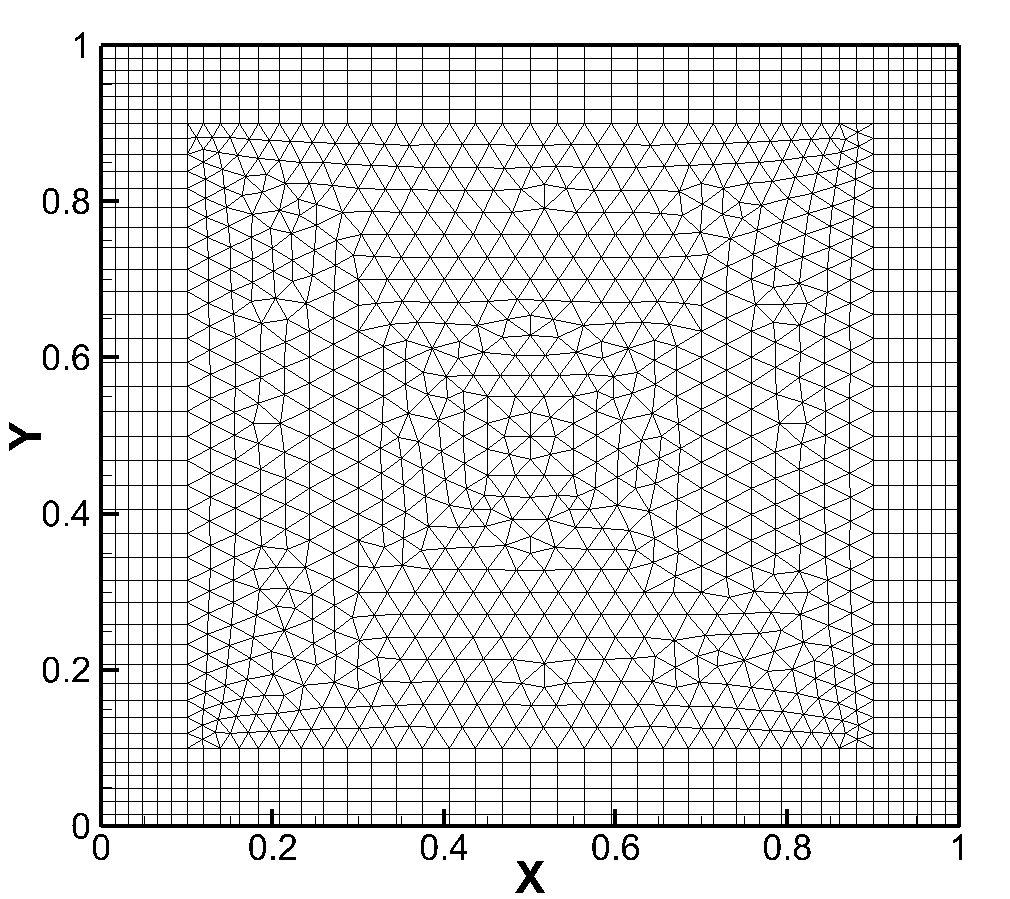}}
	\subfigure[Unstructured velocity mesh]{\label{fig:fig10b}\includegraphics[width=0.47\textwidth]{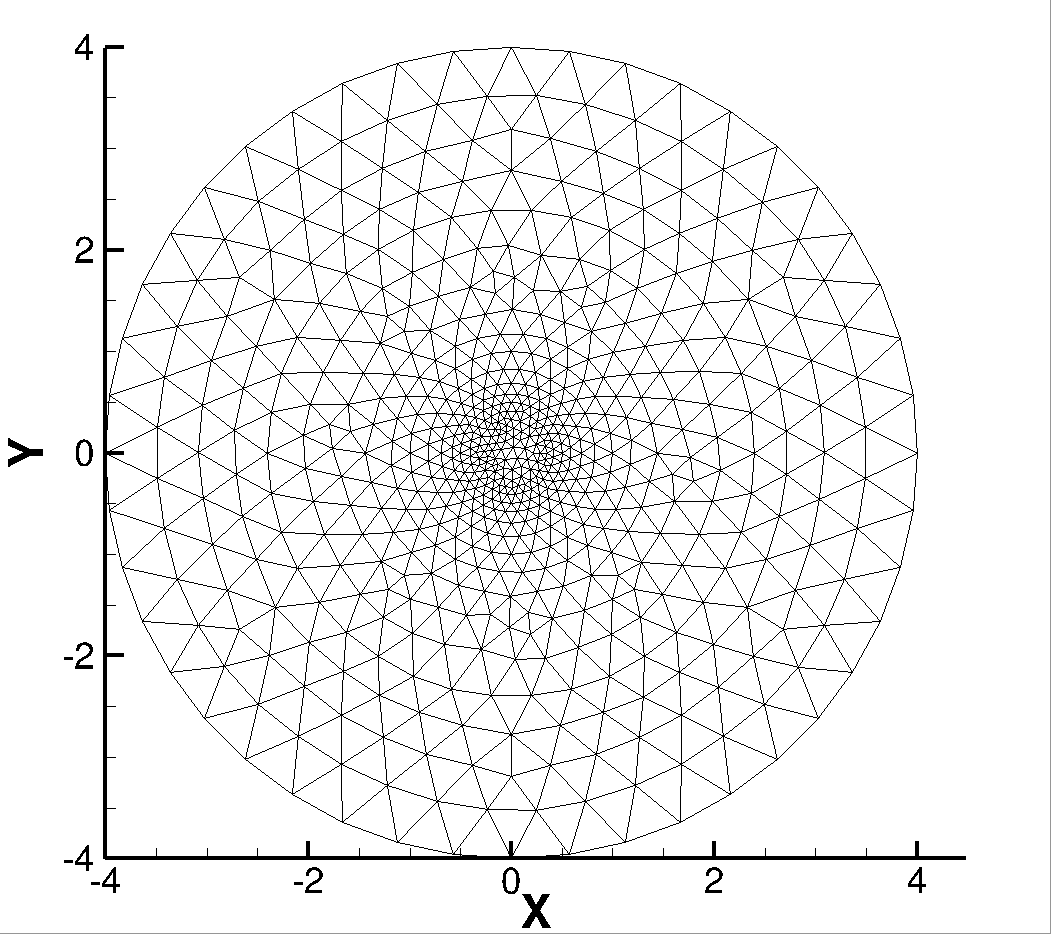}}
	\caption{\label{fig:fig10}Spatial mesh and unstructured velocity mesh for cavity flow.}
\end{figure}

Two-dimensional lid-driven cavity flow is one of the classic cases to test numerical scheme, especially their performance on viscous effect. Here, cavity flow cases with wide range of Kn numbers (0.075, 1 and 10) are carried out. The length of the four walls of the cavity (shown in Fig. \ref{fig:fig09}) are L. The top wall is moving from left to right with a constant velocity $U_w$, and the other walls are static. The working gas is Argon, and the HS molecular model is chosen. The temperatures of all walls are 273K. Initially, the flow field is in stationary with a temperature  of 273K. The sonic speed and Mach number of the moving wall are $a=\sqrt{\gamma RT}=307.64m/s$ and $M\text{a}=0.16$, respectively. In numerical simulation, $T_{ref}^{{}}=227.50K$ and $U_{ref}^{{}}=\sqrt{2RT_{ref}^{{}}}=307.64m/s$ are set as the reference temperature and velocity, respectively. Therefore, the dimensionless density and temperature of the initial flow field are $\rho _0^{} = 1.0$ and $T_0^{} = 1.2$ respectively. The dimensionless velocity of the top wall is $U_w = 0.16$. The present C-DUGKS with both structured and unstructured velocity space are used for simulation, and the UGKS data is used as the benchmark solution. The spatial mesh for cavity (2460 cells) is shown in  Fig. \ref{fig:fig10a}. In order to capture the boundary layer, a body-fitted rectangular mesh is used near the wall, while unstructured triangular mesh is used inside the cavity. For unstructured discrete velocity spaces, the  velocity mesh (1192 points) in Fig. \ref{fig:fig10b} is chosen for all Kn numbers, which is centered at $\left( {0,0} \right)$ with a radius of 4. For structured discrete velocity spaces, a structured and uniform mesh of 50x50 points is chosen for Kn=0.075, while the mesh with 101x101 points is chosen for Kn=1 and 10. The ranges of both structured velocity space are $\left[ { - 4,\;4} \right] \times \left[ { - 4,\;4} \right]$.

\begin{figure}
	\centering
	\subfigure[Temperature]{\label{fig:fig11a}\includegraphics[width=0.47\textwidth]{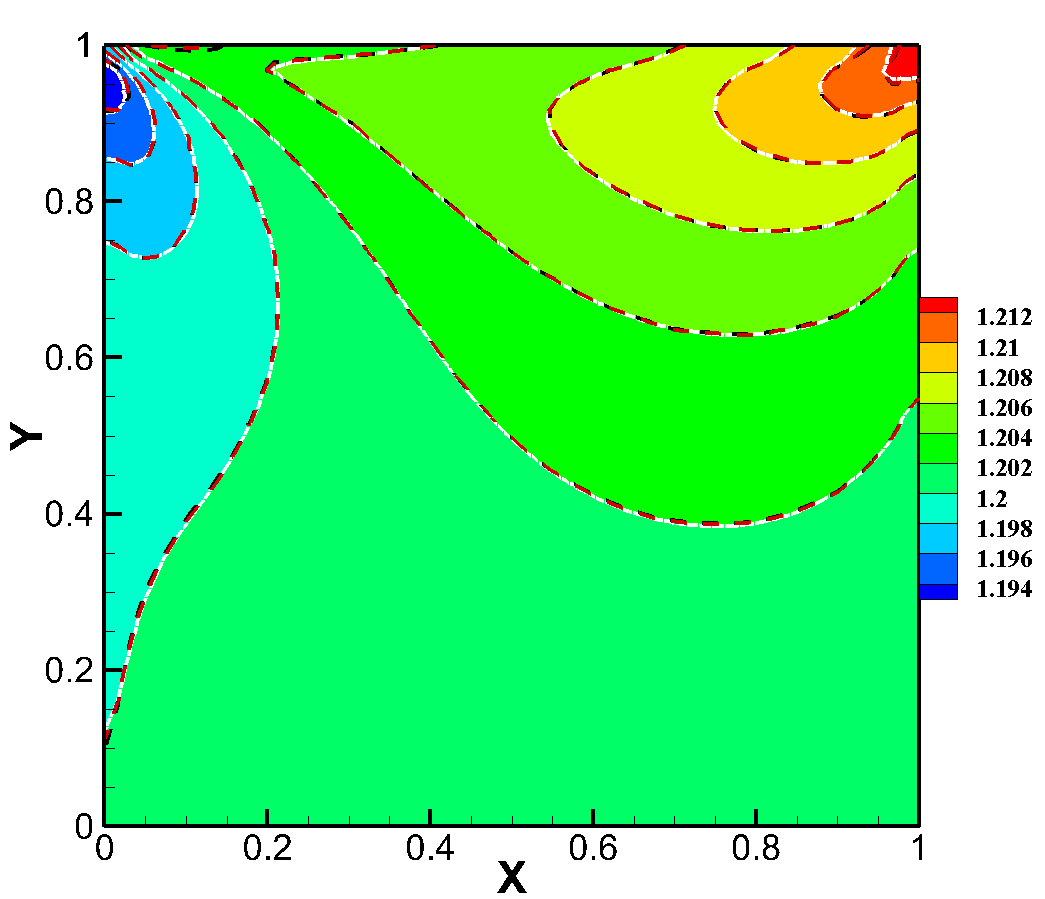}}
	\subfigure[Heat flux]{\label{fig:fig11b}\includegraphics[width=0.47\textwidth]{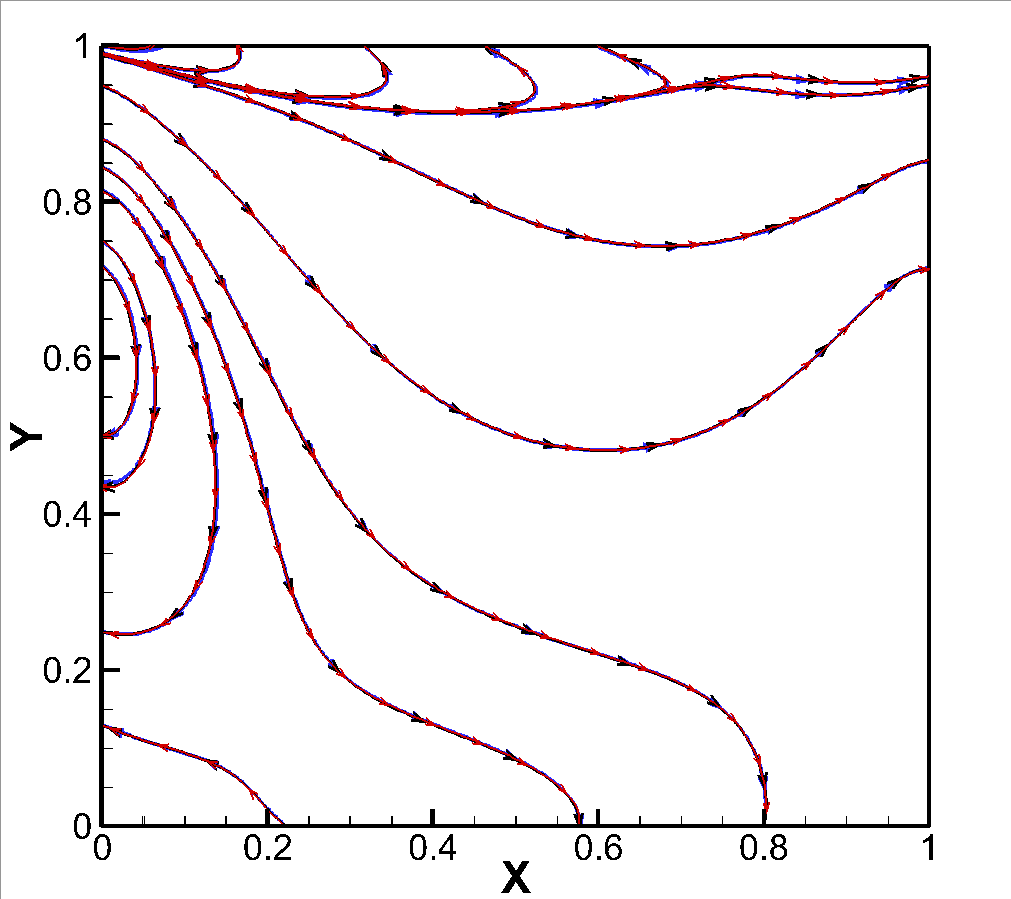}}
	\subfigure[Velocity]{\label{fig:fig11c}\includegraphics[width=0.47\textwidth]{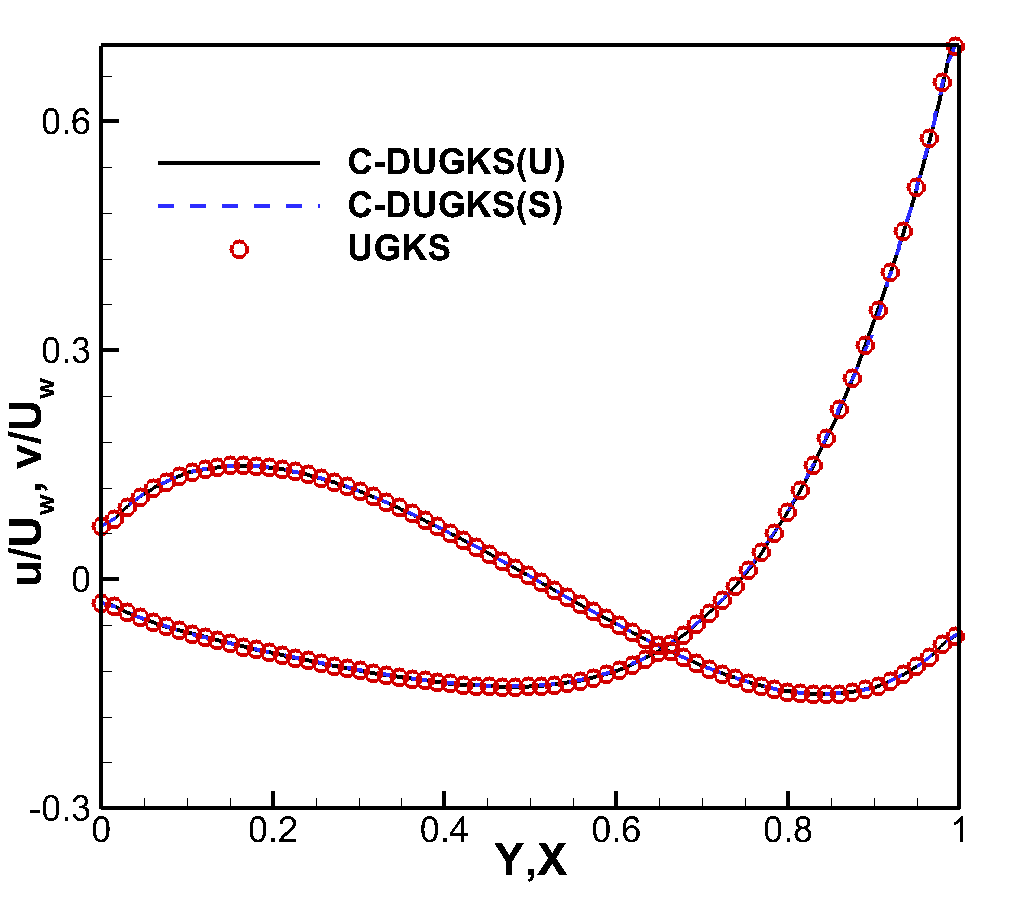}}
	\caption{\label{fig:fig11}Simulation results of cavity flow at $Kn = 0.075$. (a) Temperature (UGKS-white line with background, C-DUGKS with unstructured-black dashed line, C-DUGKS with structured-red dashed line). (b) Heat flux (UGKS-blue solid line; C-DUGKS with unstructured mesh-black solid line; C-DUGKS with structured line-red solid line). (c) U-velocity along the central vertical line and v-velocity along the central horizontal line (C-DUGKS with unstructured velocity mesh (C-DUGKS(U))-black solid line, C-DUGKS with structured velocity mesh (C-DUGKS(S))-blue dash line, UGKS-red  circle symbol).}
\end{figure}

\begin{figure}
	\centering
	\subfigure[Temperature]{\label{fig:fig12a}\includegraphics[width=0.47\textwidth]{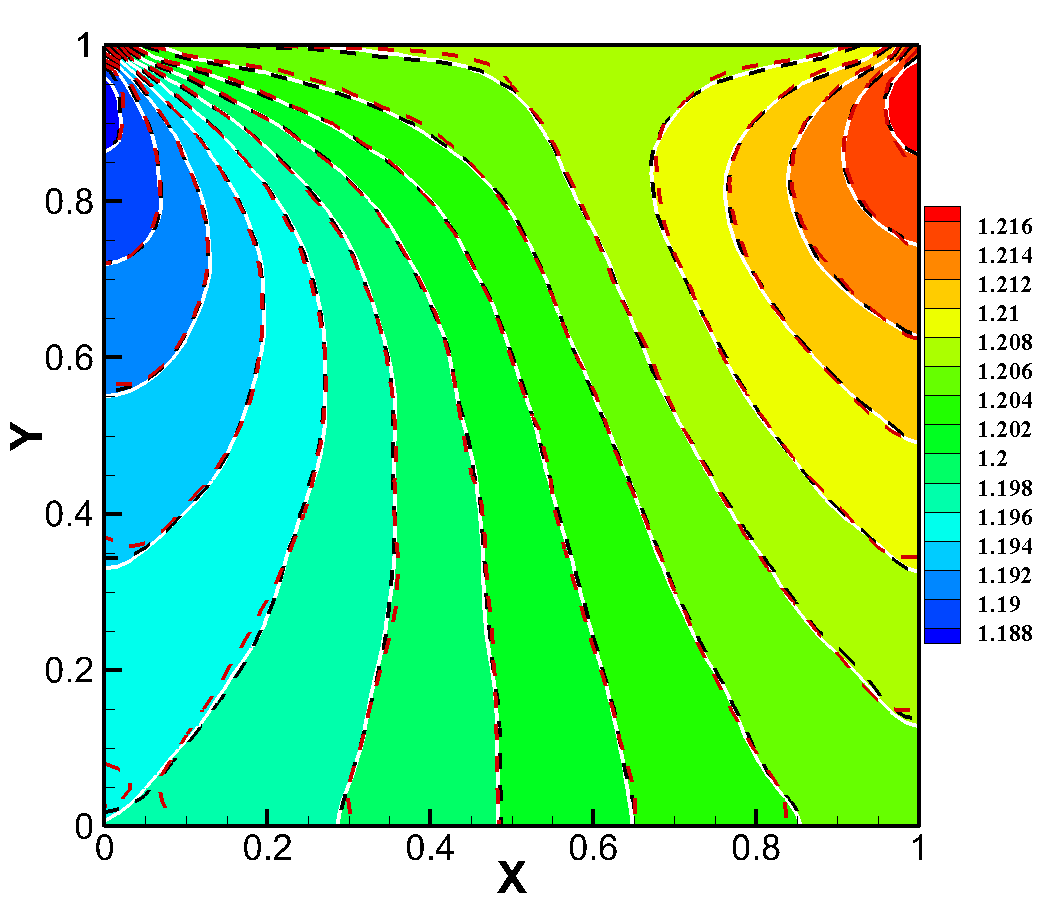}}
	\subfigure[Heat flux]{\label{fig:fig12b}\includegraphics[width=0.47\textwidth]{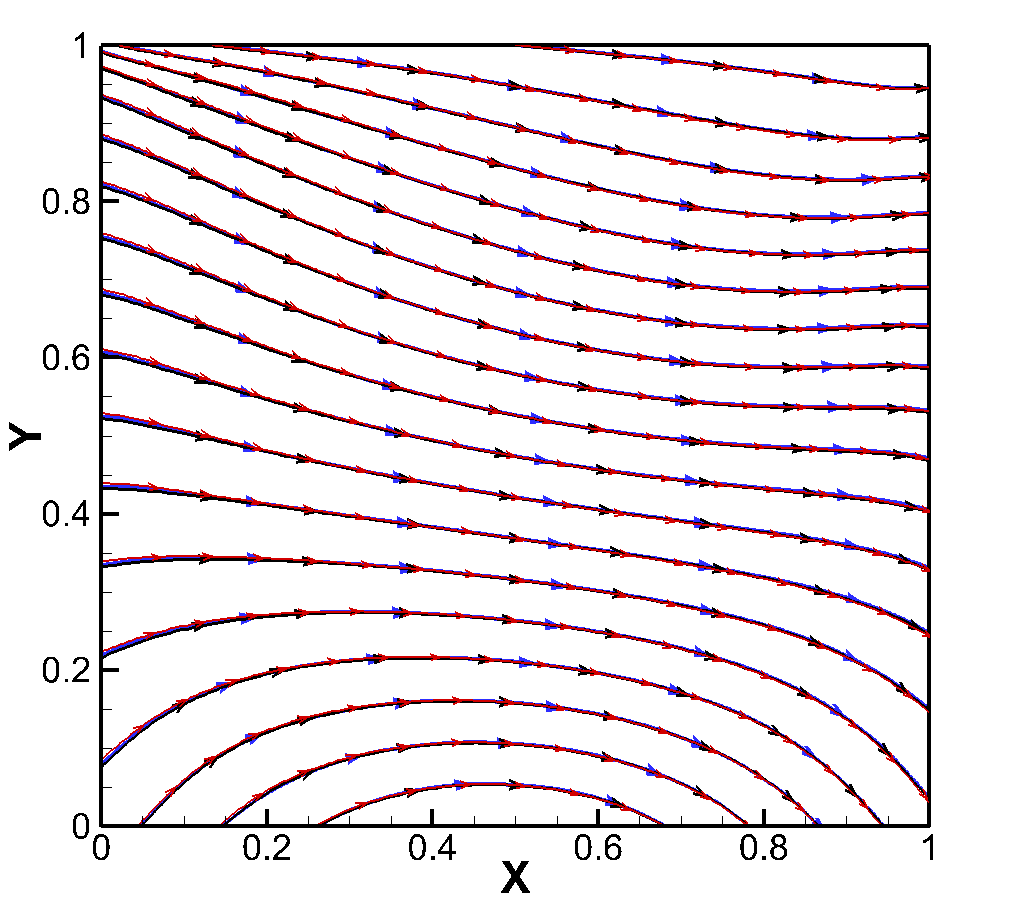}}
	\subfigure[Velocity]{\label{fig:fig12c}\includegraphics[width=0.47\textwidth]{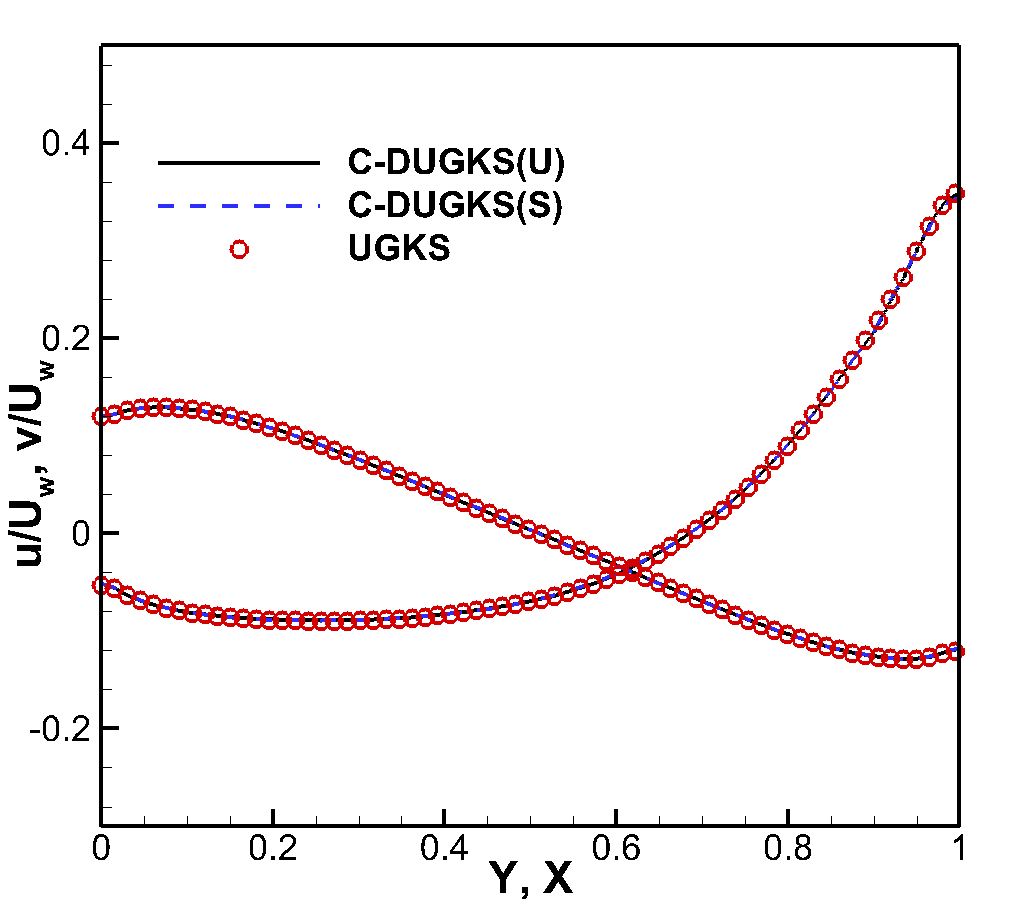}}
	\caption{\label{fig:fig12}Simulation results of cavity flow at $Kn = 10.0$. (a) Temperature (UGKS-white line with background, C-DUGKS with unstructured-black dashed line, C-DUGKS with structured-red dashed line). (b) Heat flux (UGKS-blue solid line; C-DUGKS with unstructured mesh-black solid line; C-DUGKS with structured line-red solid line). (c) U-velocity along the central vertical line and v-velocity along the central horizontal line (C-DUGKS with unstructured velocity mesh (C-DUGKS(U))-black solid line, C-DUGKS with structured velocity mesh (C-DUGKS(S))-blue dash line, UGKS-red  circle symbol).}
\end{figure}

The results of cavity flow at Kn = 0.075 and 10 are presented in Figs. \ref{fig:fig11} and \ref{fig:fig12}, where the temperature contours, the streamline of heat flux, and the velocities along the central horizontal and central vertical lines are illustrated. It can be seen that the results from C-DUGKS with both structured and unstructured velocity space are in good agreement with the UGKS. In addition, a close observation on Fig. \ref{fig:fig12a} shows that the result of C-DUGKS with the structured mesh (10201 discrete velocity points) has a nonphysical ray effect at the bottom-left corner of the cavity, i.e. the temperature contour is jittery and discontinuous. On the contrary, the result of C-DUGKS with the unstructured velocity mesh (only 1192 discrete velocity points) is consistent with the benchmark solution with no ray effect. Since the speed of the top moving wall is only 0.16 (near the origin of the velocity space), and the temperature is around 1.0 in the whole flow field, then particles are concentrated near the origin of the velocity space. Since the unstructured mesh is flexible and can arrange sufficient discrete velocity points near the velocity origin, then the distribution functions is well captured, and the ray effect is avoided. Moreover, the unstructured velocity mesh can be set coarse in the regions far from the velocity origin, then the total number of discrete velocity points is still much less than that of the structured and uniform velocity mesh. Table \ref{tab:Cavity1} and table \ref{tab:Cavity2} show the convergence time taken by C-DUGKS  with both structured and unstructured mesh. It  is found that the convergence time of C-DUGKS with the unstructured mesh is much less than that of C-DUGKS with the structured velocity space (with a speedup of 8 to 12), benefiting from use of the unstructured velocity mesh.

\begin{table}\label{tab:Cavity1}
	\centering
	\caption{Convergence time of C-DUGKS with both structured and unstructured velocity meshes (in seconds) for Kn=0.075 case of Cavity flow.}
	\begin{tabular}{p{132pt} p{60pt}<{\raggedleft} p{60pt}<{\raggedleft} p{30pt}<{\raggedleft}  }
		\hline
		\hline
		& Structured mesh (A) & Unstructured mesh (B) & ratio (A/B)\\
		\hline
		Discrete velocity points & 2500 & 1192 & 2.1\\
		\hline
		Convergence time (Kn=0.075) &  677.1 & 235.7 & 2.9 \\
		\hline
		\hline
	\end{tabular}
\end{table}

\begin{table}\label{tab:Cavity2}
	\centering
	\caption{Convergence time of C-DUGKS with both structured and unstructured velocity meshes (in seconds) for Kn=1 and 10 cases of Cavity flow.}
	\begin{tabular}{p{130pt} p{60pt}<{\raggedleft} p{60pt}<{\raggedleft} p{30pt}<{\raggedleft}  }
		\hline
		\hline
		& Structured mesh (A) & Unstructured mesh (B) & ratio (A/B)\\
		\hline
		Discrete velocity points & 10201 & 1192 & 8.6\\
		\hline
		Convergence time (Kn=1) &  1537.3 & 189.3 & 8.1 \\
		\hline
		Convergence time (Kn=10) & 2319.1 & 189.1 & 12.3 \\
		\hline
		\hline
	\end{tabular}
\end{table}

\subsection{Two-dimensional rarefied Riemann problem}\label{2D Riemann problem}
In this section, the two-dimensional rarefied Riemann problem is used to test the validity of the present C-DUGKS method. And the computational domain is divided into quadrants with different macroscopic variables as follows:
\begin{equation}\label{equ41}
\left( {\rho ,u,v,p} \right) = \left\{ {\begin{array}{*{20}{c}}
	{\begin{array}{*{20}{c}}
		{\left( {{\rho _1},{u_1},{v_1},{p_1}} \right) = \left( {0.5313,0,0,0.4} \right)\;\;\;\;x > 0,\;y > 0,} \\
		{\left( {{\rho _2},{u_2},{v_2},{p_2}} \right) = \left( {1,0.7276,0,1} \right)\;\;\;\;\;\;x \leqslant 0,\;y > 0,}
		\end{array}} \\
	{\begin{array}{*{20}{c}}
		{\left( {{\rho _3},{u_3},{v_3},{p_3}} \right) = \left( {0.8,0,0,1} \right)\;\;\;\;\;\;\;\;\;\;\;\;x \leqslant 0,\;y \leqslant 0,} \\
		{\left( {{\rho _4},{u_4},{v_4},{p_4}} \right) = \left( {1,0,0.7276,1} \right)\;\;\;\;\;\;x > 0,\;y \leqslant 0.}
		\end{array}}
	\end{array}} \right.
\end{equation}

\begin{figure}
	\centering
	\includegraphics[width=0.6\textwidth]{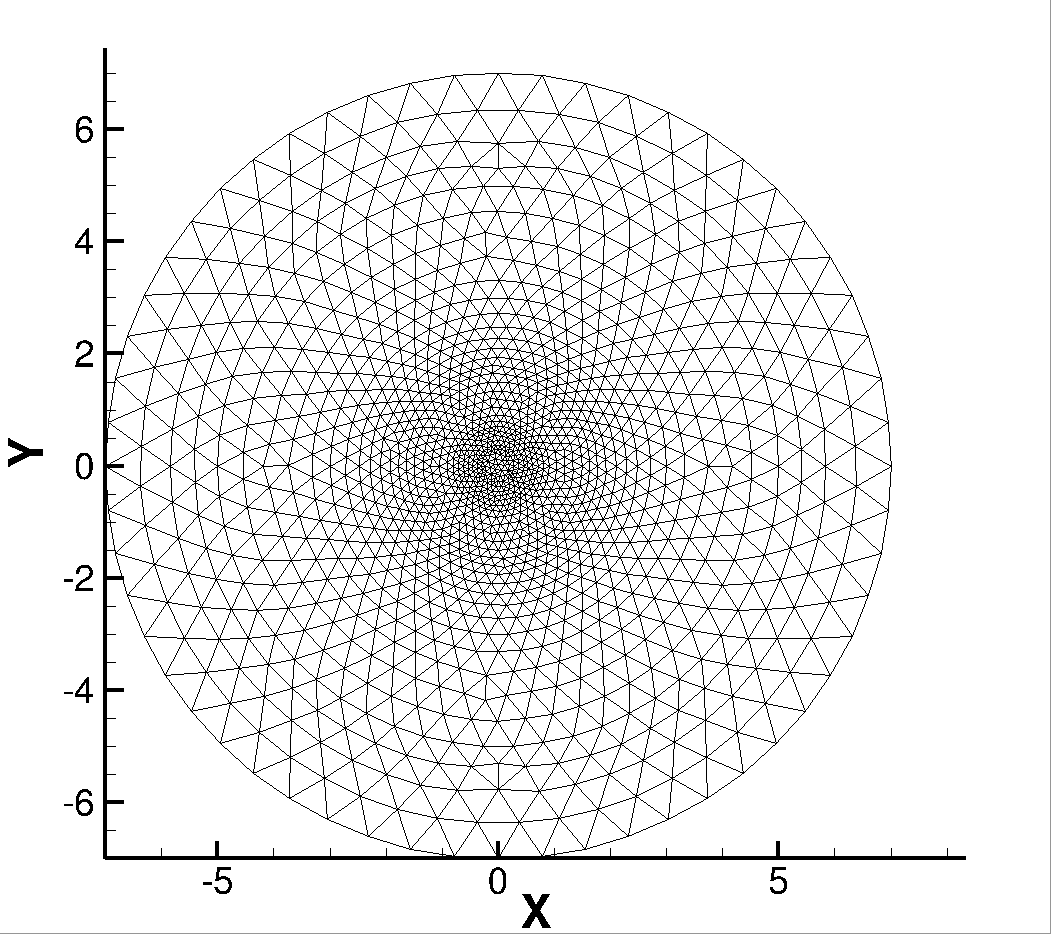}
	\caption{\label{fig:fig13} Unstructured velocity mesh for 2D rarefied Riemann problem.}
\end{figure}

The working gas is composed by diatomic molecule (HS model) with $\gamma  = 1.4$. The viscous coefficient is $\mu _0^{} = 10$. Taking the physical variables in the first quadrant as the reference values, then Kn is 27.7. The computational domain $\left[ - {{\text{ 0.5 }},\;0.5} \right] \times \left[-{{\text{ 0.5 }},\;0.5} \right]$ is discretized into a Cartesian mesh with 60x60 cells. The unstructured discrete velocity spaces used in C-DUGKS (2788 discrete velocity points) is shown in Fig. \ref{fig:fig13}, which is centered at $\left( {0,0} \right)$ with a radius of 7. While the structured discrete velocity spaces used in C-DUGKS has 101x101 discrete velocity points in the range of  $\left[ { - {\text{7}},\;{\text{7}}} \right] \times \left[ { - {\text{7}},\;{\text{7}}} \right]$.

\begin{figure}
	\centering
	\subfigure[Density]{\label{fig:fig14a}\includegraphics[width=0.47\textwidth]{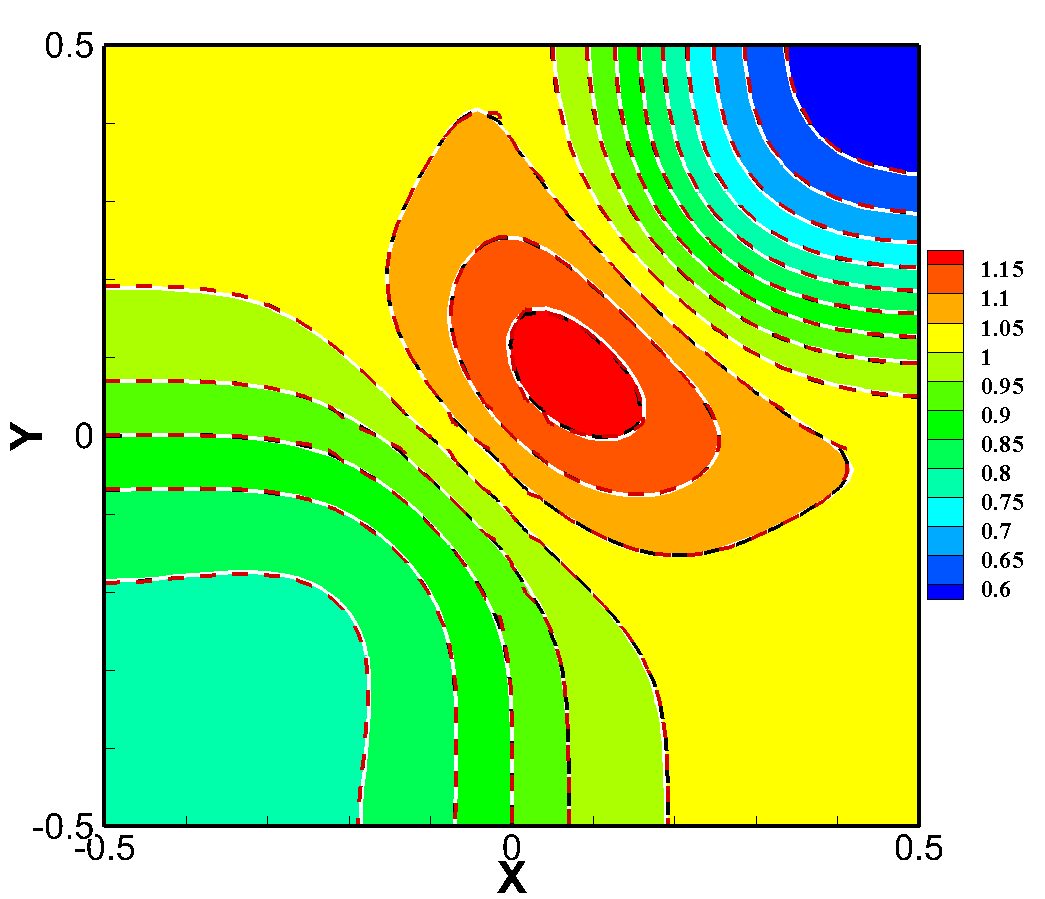}}
	\subfigure[Temperature]{\label{fig:fig145}\includegraphics[width=0.47\textwidth]{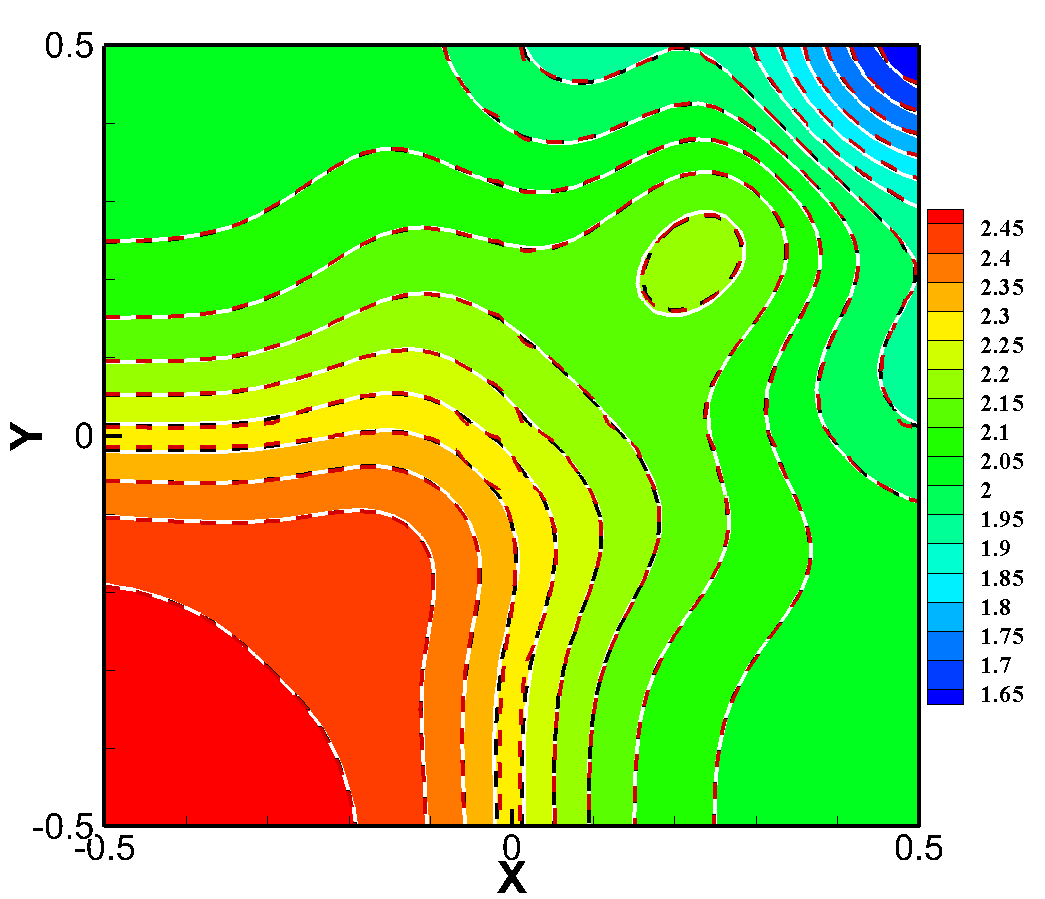}}
	\subfigure[Velocity magnitude]{\label{fig:fig14c}\includegraphics[width=0.47\textwidth]{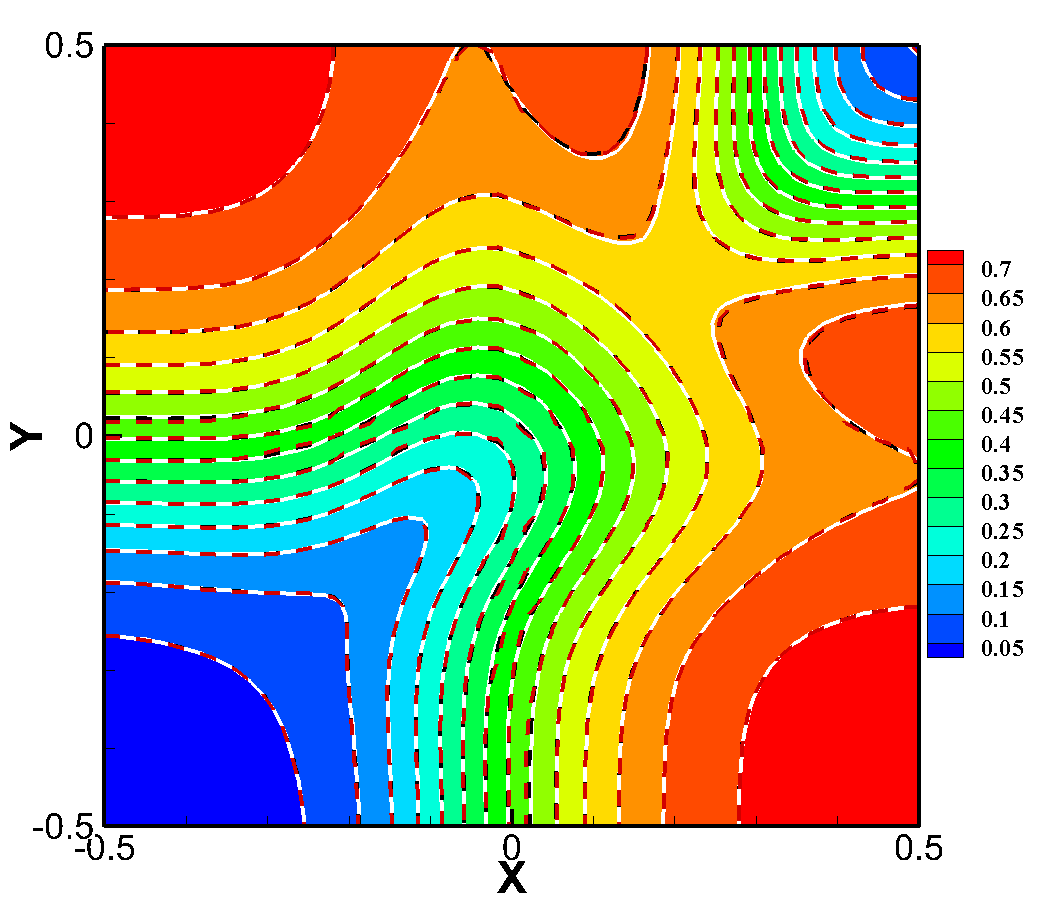}}
	\subfigure[Streamlines]{\label{fig:fig14d}\includegraphics[width=0.47\textwidth]{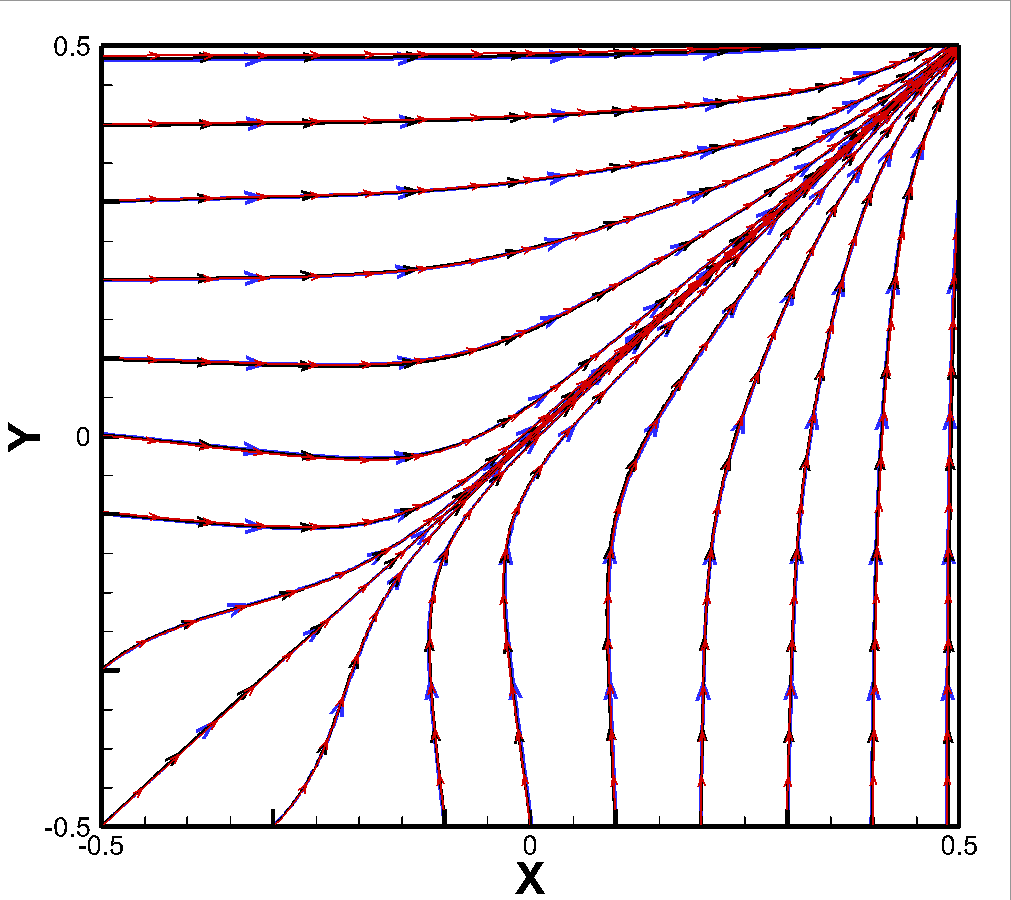}}
	\caption{\label{fig:fig14}The contours of the density, temperature and velocity magnitude and the streamlines  for 2D rarefied Riemann problem at t=0.15. In subfigure (a), (b) and (c), the background and white solid lines are from the collisionless Boltzmann equation, the black dash lines and red dash lines are the results of C-DUGKS witn unstructured and structured velocity mesh, respectively. In subfigure (d), the blue solid lines are the solutions of the collisionless Boltzmann equation, the black solid lines and red solid lines are the results of C-DUGKS witn unstructured and structured velocity mesh, respectively.}
\end{figure}

The density, temperature, velocity magnitude $ \sqrt{ u_x^2 + u_y^2 }  $, and streamlines at t = 0.15 are shown in Fig. \ref{fig:fig14}. The results of C-DUGKS with both structured and unstructured mesh match well with the results from the solution of the collisionless Boltzmann equation. Table \ref{tab:Riemann} is number of discrete velocity points and the computation time of the C-DUGKS with both structured and unstructured mesh, respectively. It is found that the speed-up ratio is the same with the ratio of the mesh numbers (the number of structured velocity mesh over the number of unstructured velocity mesh). Although the velocity space is unstructured, the algorithm of C-DUGKS only use the information of the coordinates of the discrete velocity points (the centers of the mesh cells) and their weights (the areas (2-D) of the mesh cells), other information, such as the face-node connection, face-cell connection, etc. are not used. As a results, the efficiencies of structured and unstructured mesh are almost the same if used as the discrete velocity space. Moreover, since the unstructured mesh is more flexible, the number of discrete velocity points can be reduced significantly, leading to a large speed-up ratio.

\begin{table}\label{tab:Riemann}
	\centering
	\caption{Run time of C-DUGKS with both structured and unstructured velocity meshes (in seconds) for Riemann problem.}
	\begin{tabular}{p{101pt} p{60pt}<{\raggedleft} p{60pt}<{\raggedleft} p{30pt}<{\raggedleft}  }
		\hline
		\hline
		& Structured mesh (A) & Unstructured mesh (B) & ratio (A/B)\\
		\hline
		Discrete velocity points & 10201 & 2788 & 3.7\\
		\hline
		Run time (Kn=27.7) &  642.2 & 175.9 & 3.7 \\
		\hline
		\hline
	\end{tabular}
\end{table}

\subsection{Supersonic flow over a cylinder}\label{cylinder}
\begin{figure}
	\centering
	\includegraphics[width=0.6\textwidth]{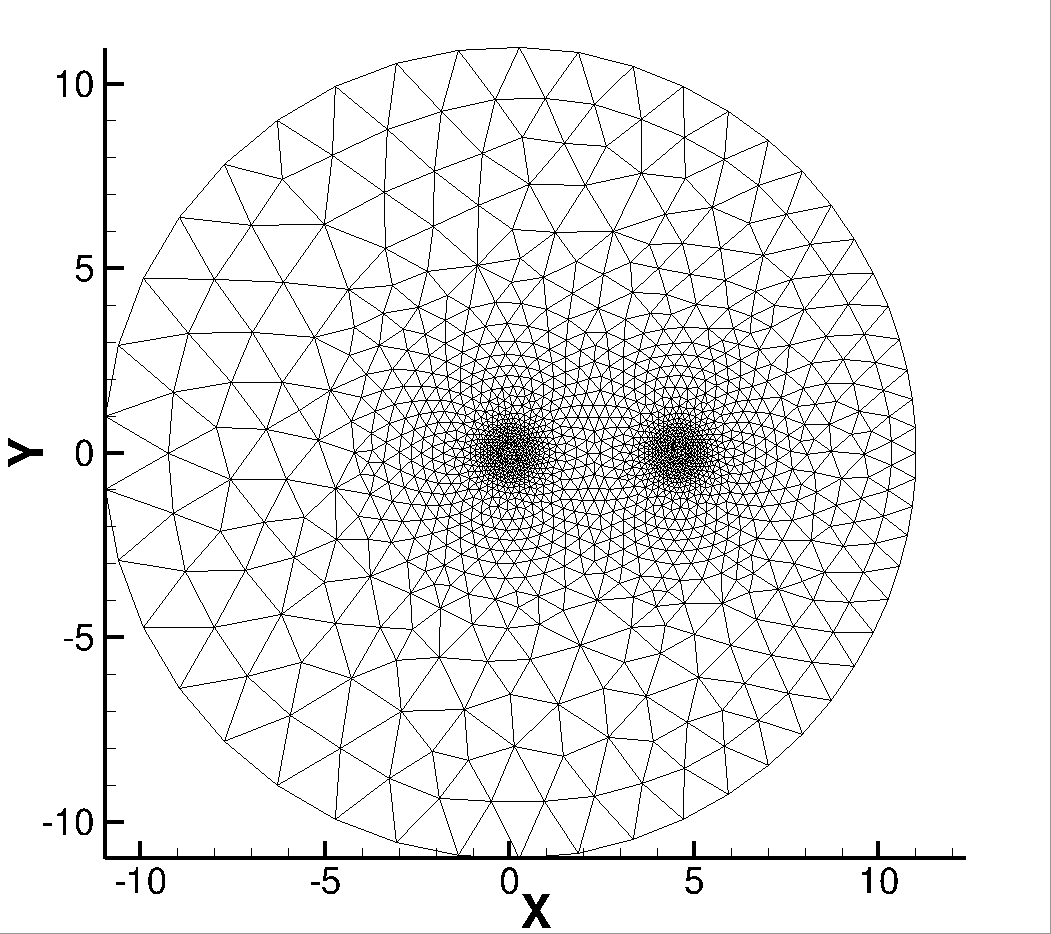}
	\caption{\label{fig:fig15} Unstructured velocity mesh for C-DUGKS.}
\end{figure}

To examine the accuracy and efficiency of the present C-DUGKS method for predicting high-speed non-equilibrium flows, the supersonic flow over a circular cylinder are conducted. The working gas is Argon for which the VHS molecular model is chosen. The velocity and temperature of the free stream are 1538.18m/s and 273K, respectively, yielding a free stream Mach number equal to 5. The temperature on the wall of cylinder is a constant at 273K. The above setting is identical to that in Ref. \cite{Zhu2016Implicit}. The radius $R$ of the cylinder is taken as the reference length. Two Kn numbers 0.1 and 1.0 in transition regime are chosen for simulation. The computational domain of physical space, which is a circle with a radius of 15R, is discretized using a mesh of 64x61 cells. In order to capture the boundary layer correctly, the mesh near the cylinder wall  is refined in the normal direction. The height of the first layer near the solid boundary is set 0.005 and 0.01 for Kn=0.1 and 1, respectively, since the thickness of boundary layer decreases as Kn number decreases. During the simulation, the reference temperature and velocity are $T_{ref}^{{}}=273K$ and $U_{ref}^{{}}=\sqrt{2RT_{ref}^{{}}}=337.00m/s$, respectively. Therefore, the dimensionless free-stream temperature and the wall temperature are both 1.0, the free-stream density is 1.0 and free-stream velocity is 4.56. The dimensionless density and temperature of the initial flow field are $\rho _0^{} = 1.0$  and $T_0^{} = 1.0$ respectively.
The unstructured velocity mesh (3045 points) used in C-DUGKS is shown in Fig. \ref{fig:fig15}, which is centered at $\left( {0,\;0} \right)$ with a radius of 10. Note that, since the free-stream velocity is 4.56, the unstructured velocity mesh is refined near both the origin $\left( {0,\;0} \right)$ and the point $\left( {4.56,\;0} \right)$. For structured velocity mesh, 89x89 (7921) points is used, and the range of discrete velocity is $\left[ { - {\text{10}},\;{\text{10}}} \right] \times \left[ { - {\text{10}},\;{\text{10}}} \right]$.

\begin{figure}
	\centering
	\subfigure[Density]{\label{fig:fig16a}\includegraphics[width=0.47\textwidth]{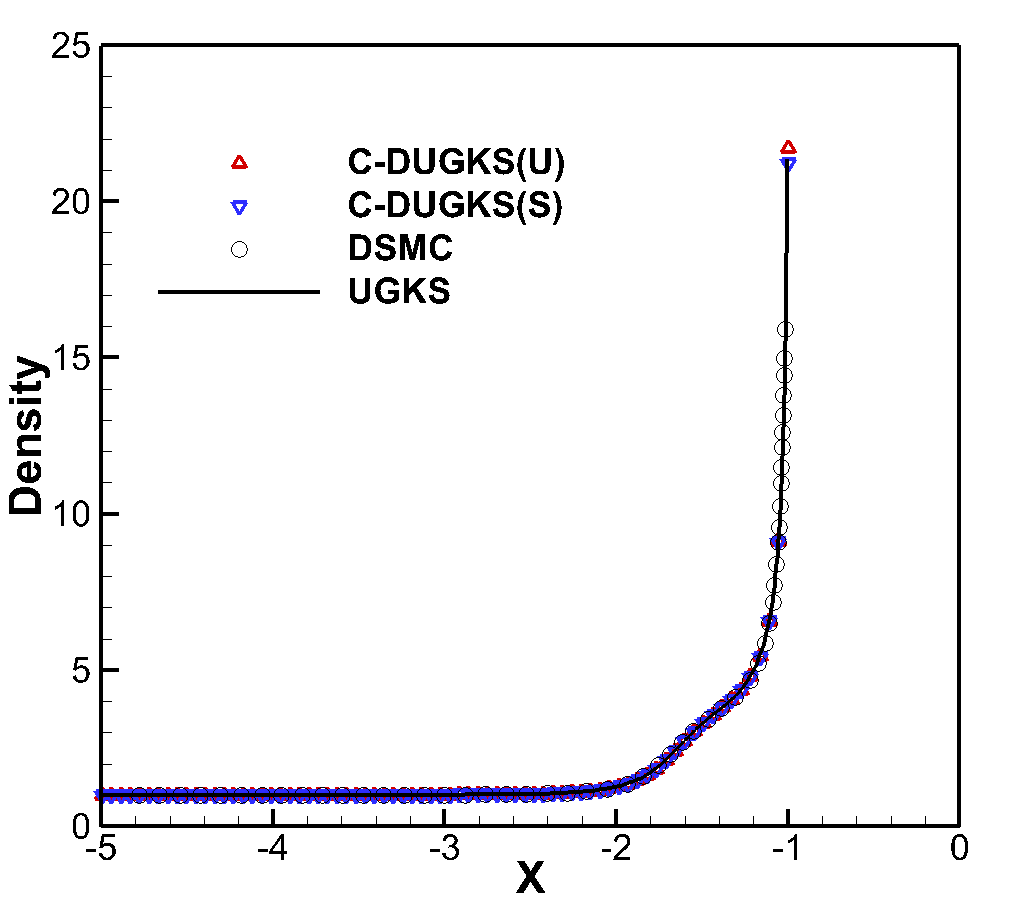}}
	\subfigure[Pressure]{\label{fig:fig16b}\includegraphics[width=0.47\textwidth]{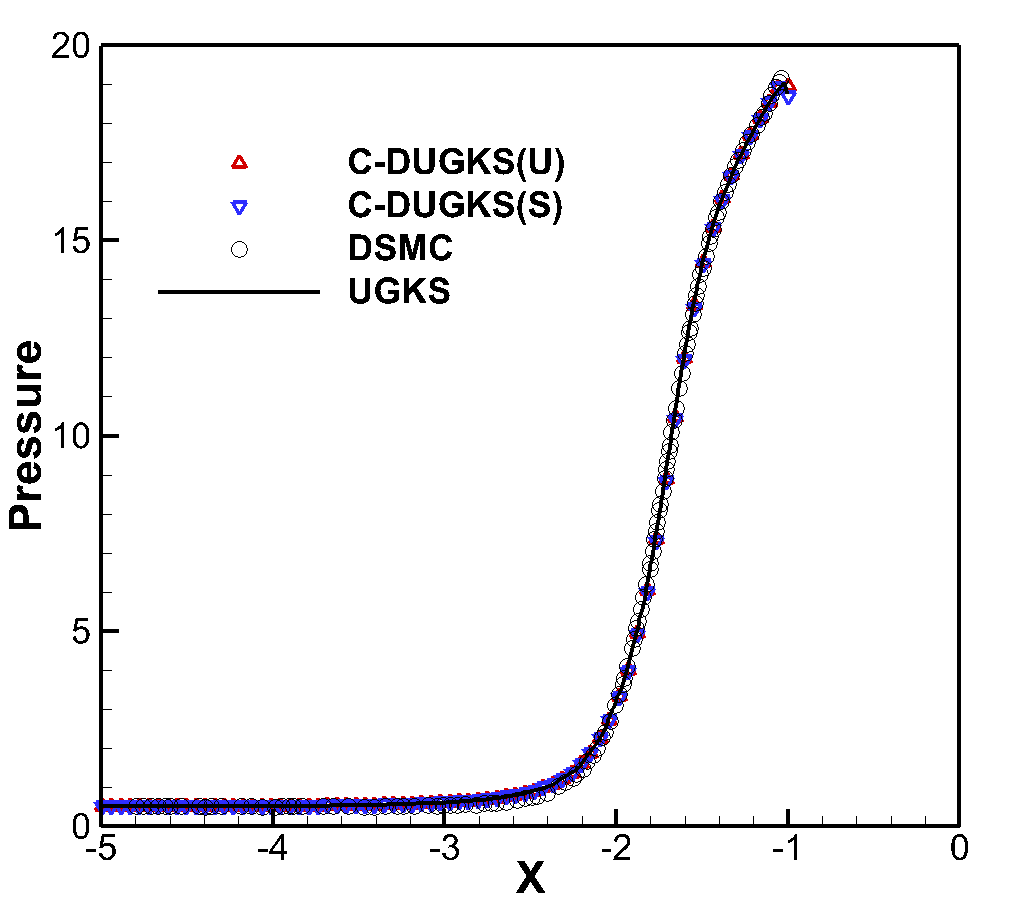}}
	\subfigure[Temperature]{\label{fig:fig16c}\includegraphics[width=0.47\textwidth]{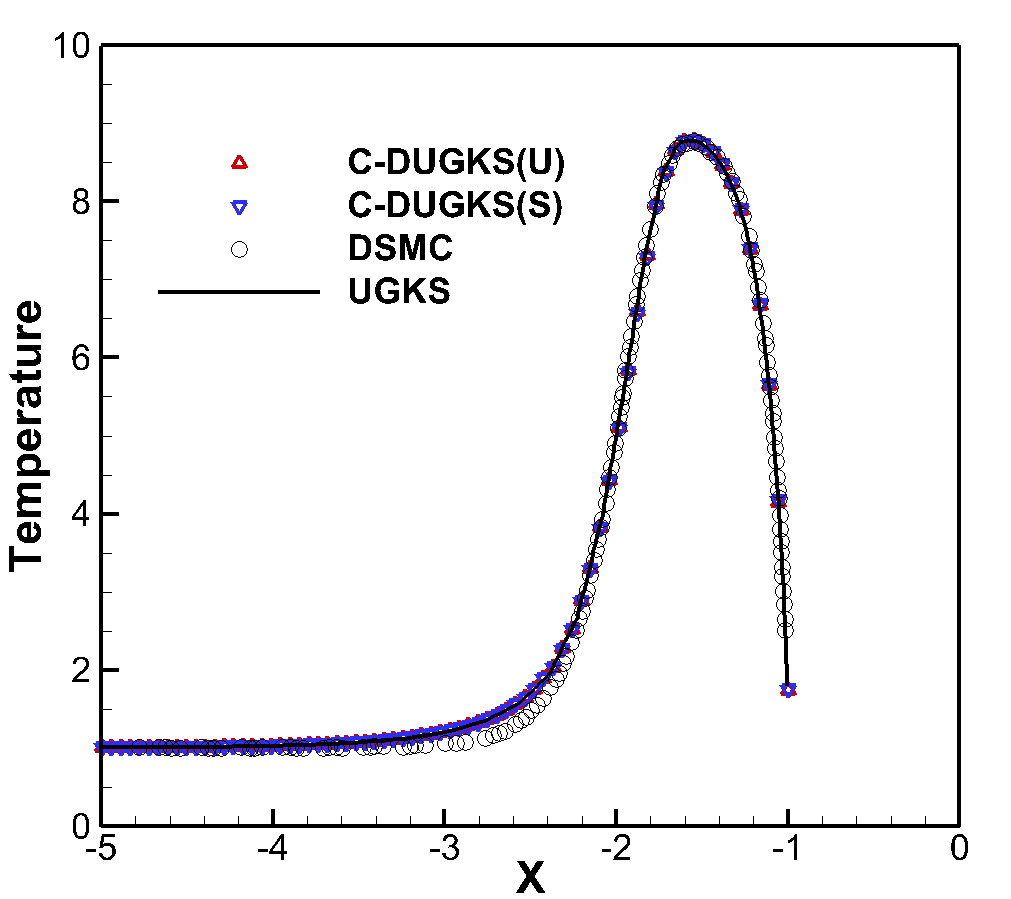}}
	\subfigure[Horizontal velocity]{\label{fig:fig16d}\includegraphics[width=0.47\textwidth]{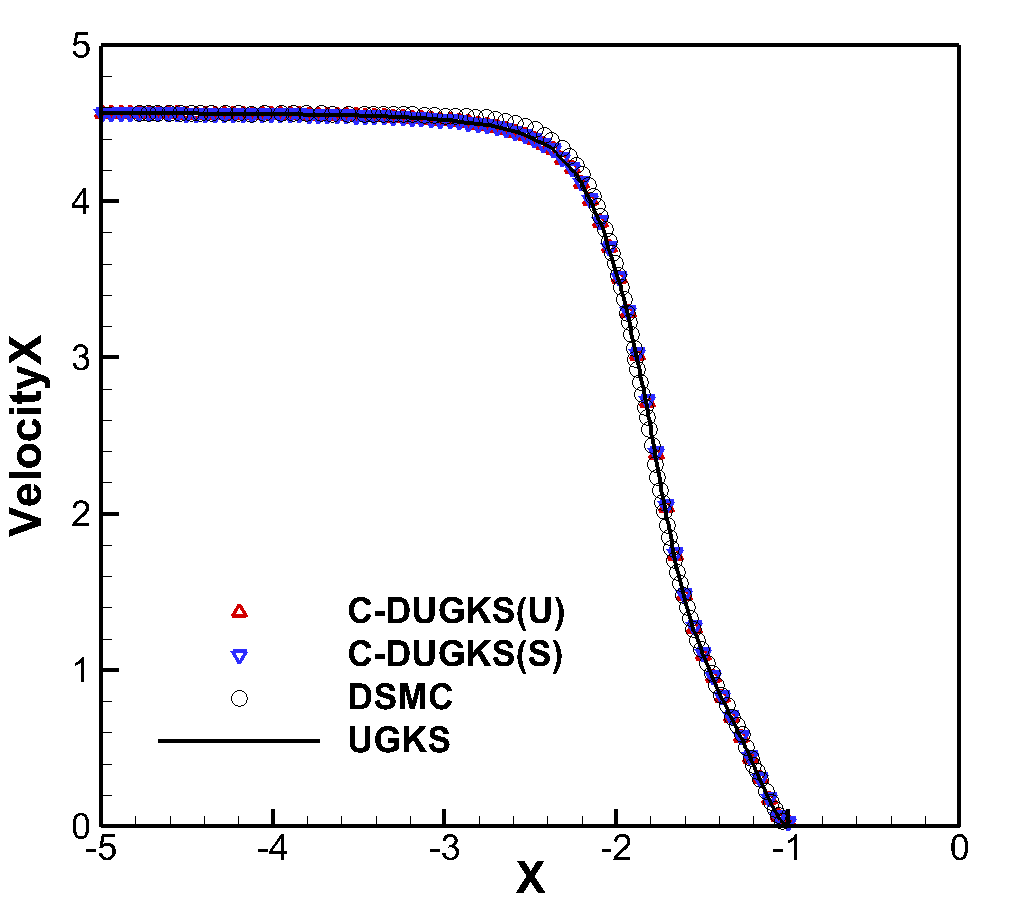}}
	\caption{\label{fig:fig16} Flow variables along the stagnation line for the case of $Kn = 0.1$ (C-DUGKS(U) and C-DUGKS(S) represent C-DUGKS with unstructured and structured velocity mesh, respectively).}
\end{figure}

\begin{figure}
	\centering
	\subfigure[Heat flux]{\label{fig:fig17a}\includegraphics[width=0.47\textwidth]{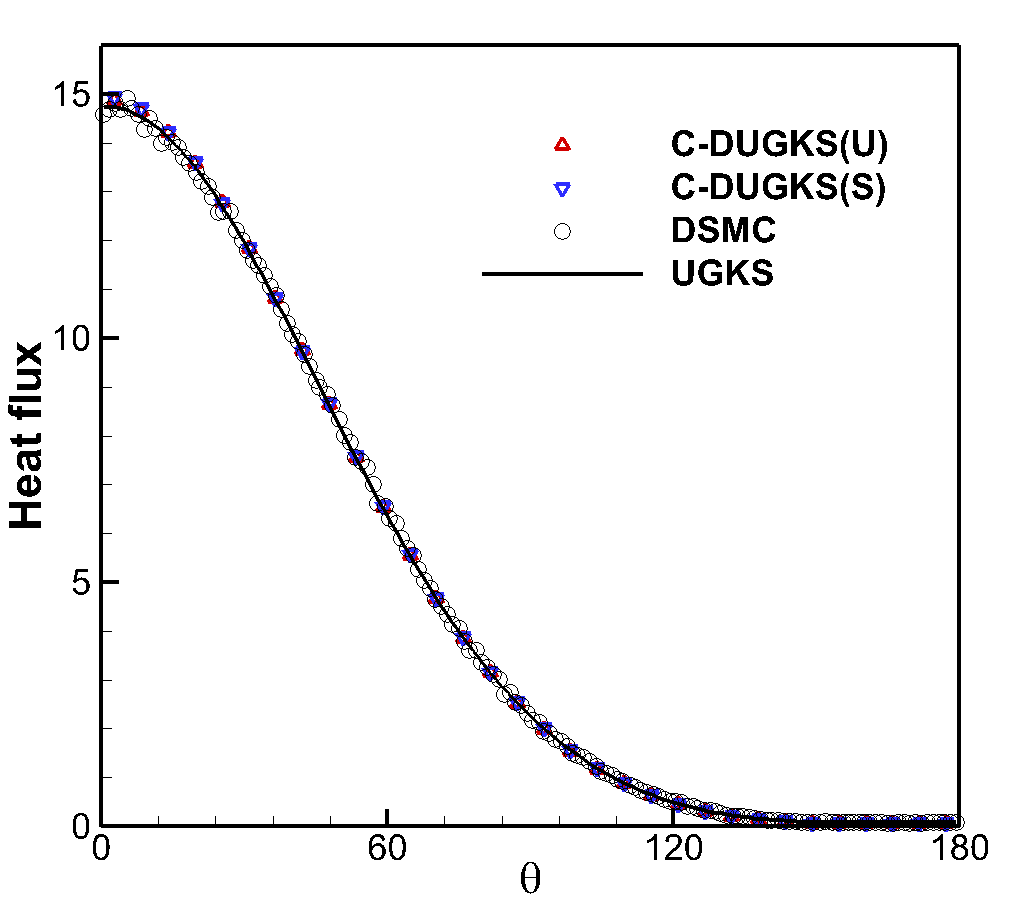}}
	\subfigure[Pressure]{\label{fig:fig17b}\includegraphics[width=0.47\textwidth]{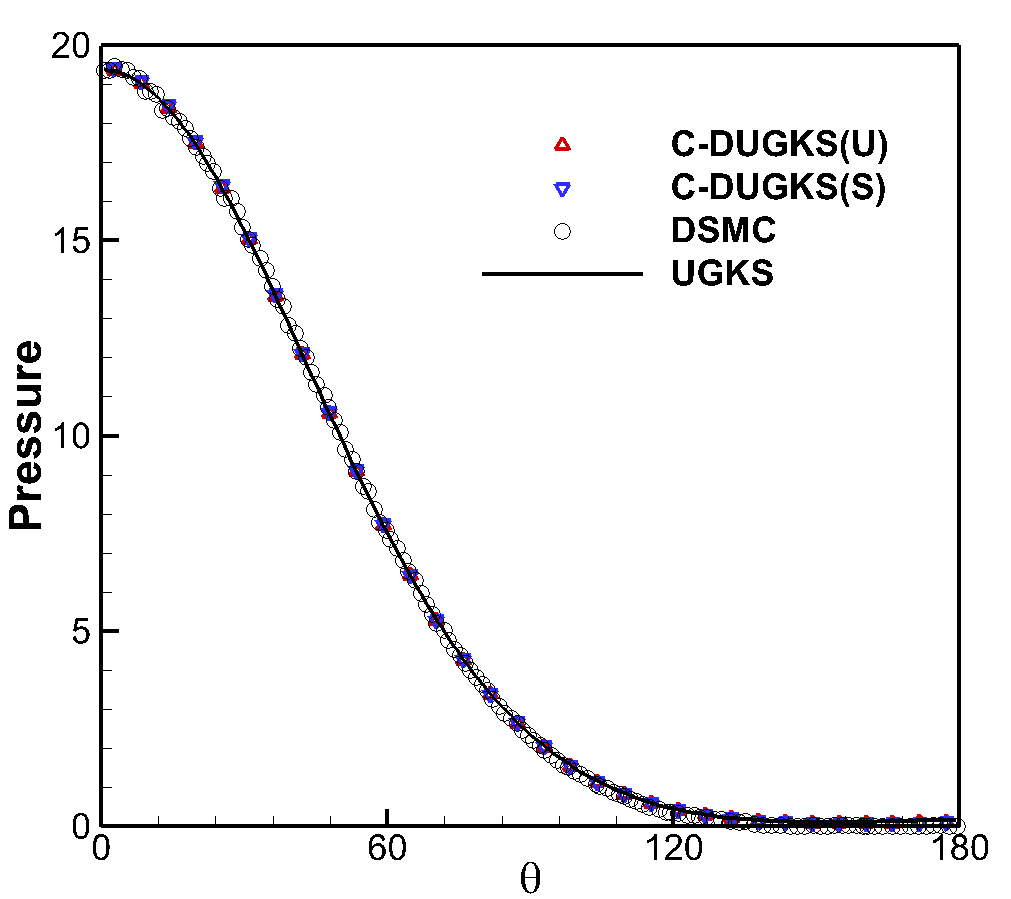}}
	\subfigure[Shear stress]{\label{fig:fig17c}\includegraphics[width=0.47\textwidth]{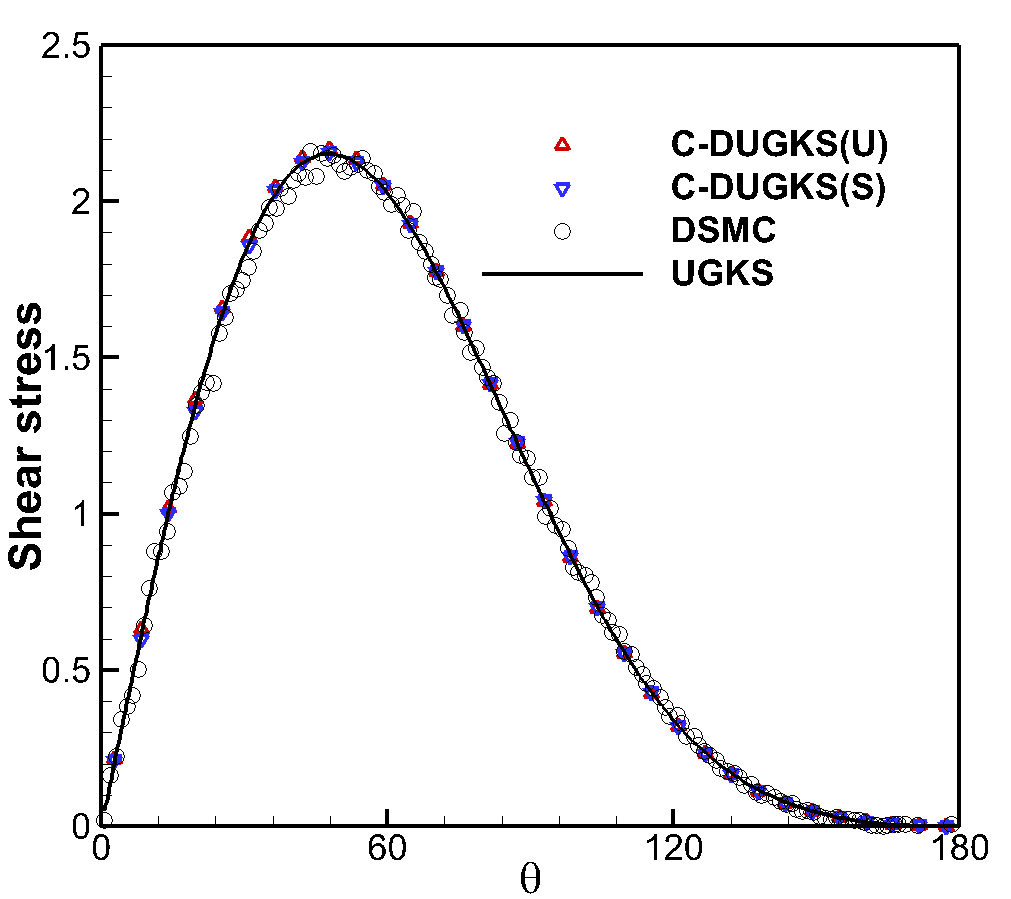}}
	\caption{\label{fig:fig17} Flow variables along the surface of the cylinder for the case of $Kn = 0.1$ (C-DUGKS(U) and C-DUGKS(S) represent C-DUGKS with unstructured and structured velocity mesh, respectively).}
\end{figure}

\begin{figure}
	\centering
	\subfigure[Density]{\label{fig:fig18a}\includegraphics[width=0.47\textwidth]{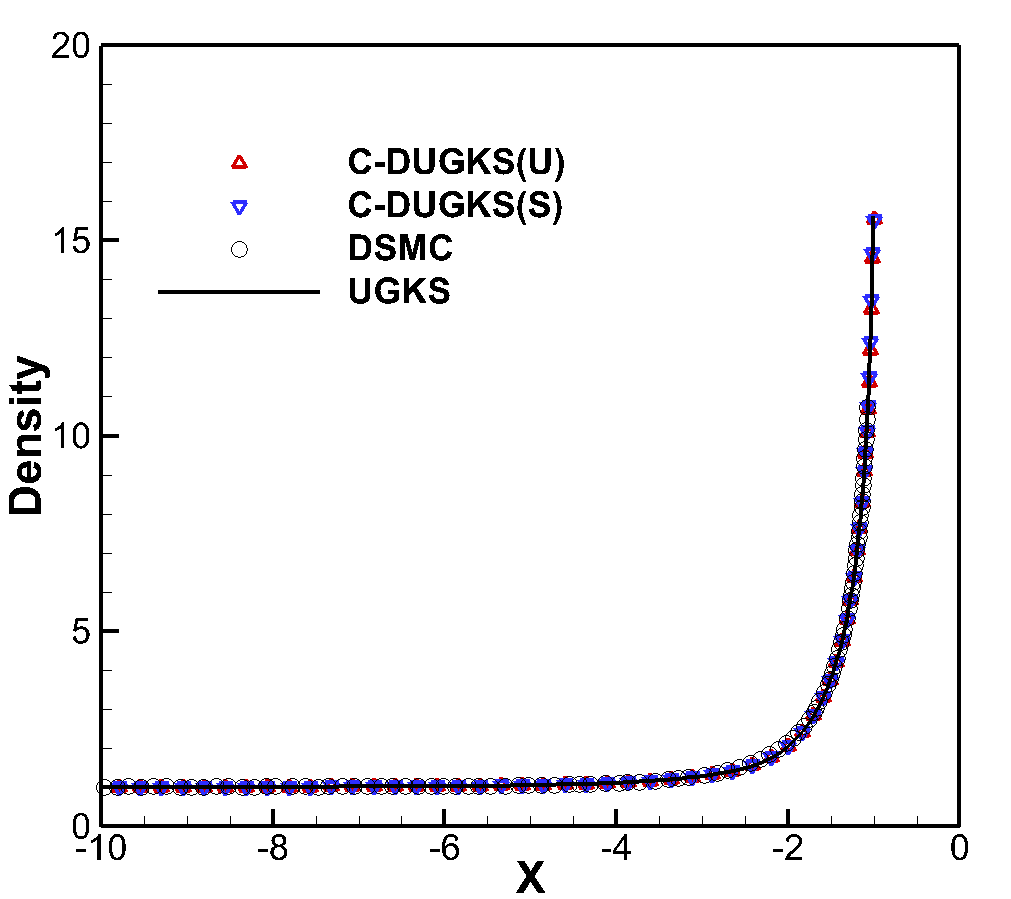}}
	\subfigure[Pressure]{\label{fig:fig18b}\includegraphics[width=0.47\textwidth]{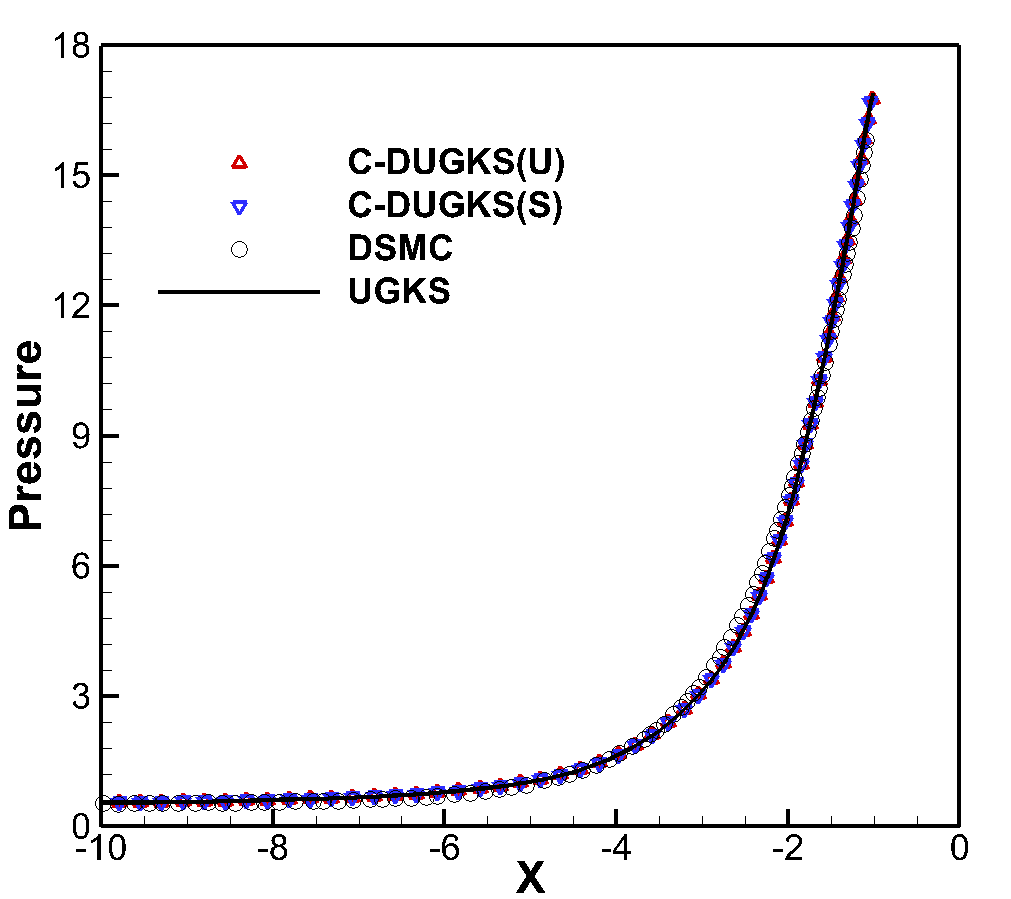}}
	\subfigure[Temperature]{\label{fig:fig18c}\includegraphics[width=0.47\textwidth]{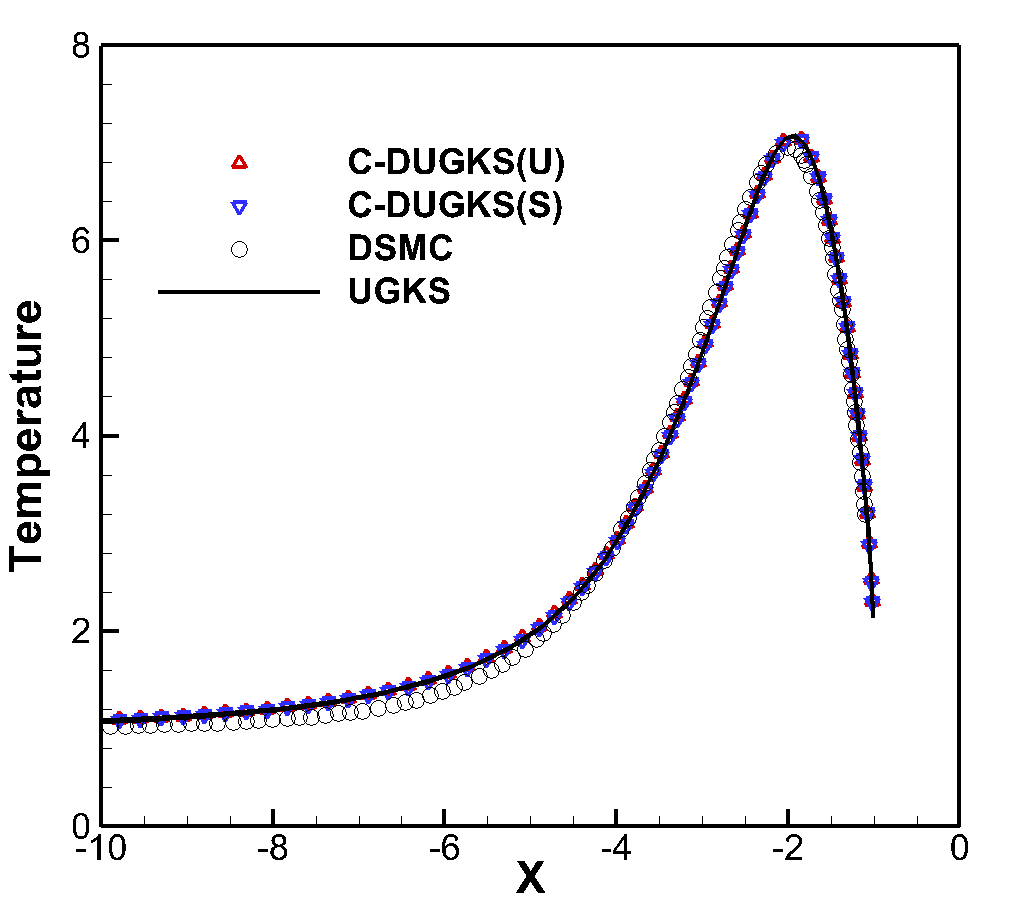}}
	\subfigure[Horizontal velocity]{\label{fig:fig18d}\includegraphics[width=0.47\textwidth]{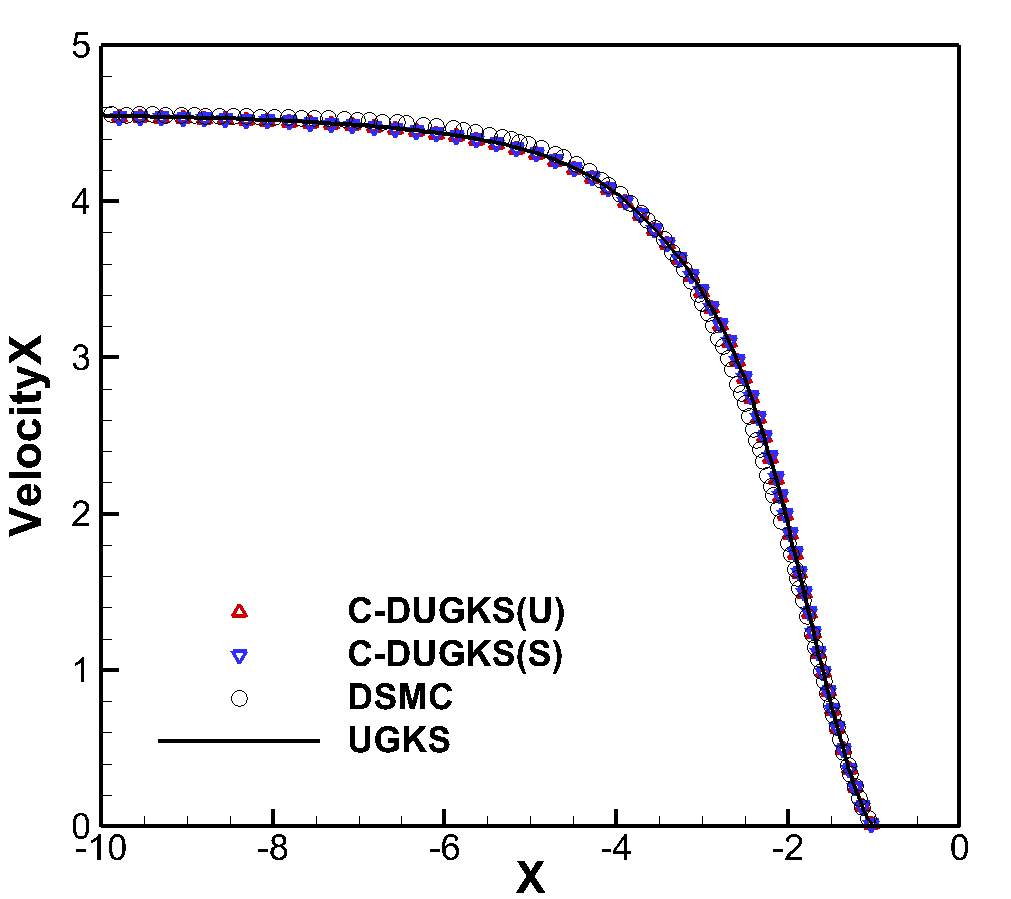}}
	\caption{\label{fig:fig18} Flow variables along the stagnation line for the case of $Kn = 1.0$ (C-DUGKS(U) and C-DUGKS(S) represent C-DUGKS with unstructured and structured velocity mesh, respectively).}
\end{figure}

\begin{figure}
	\centering
	\subfigure[Heat flux]{\label{fig:fig19a}\includegraphics[width=0.47\textwidth]{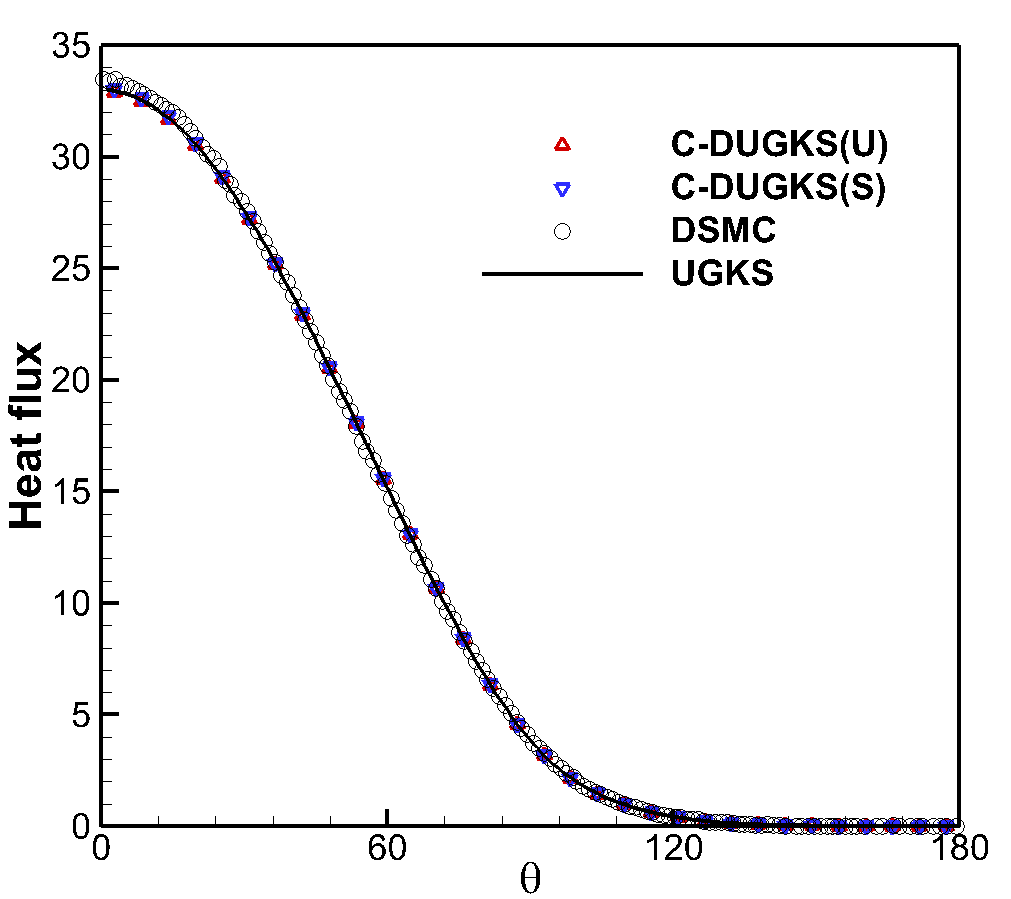}}
	\subfigure[Pressure]{\label{fig:fig19b}\includegraphics[width=0.47\textwidth]{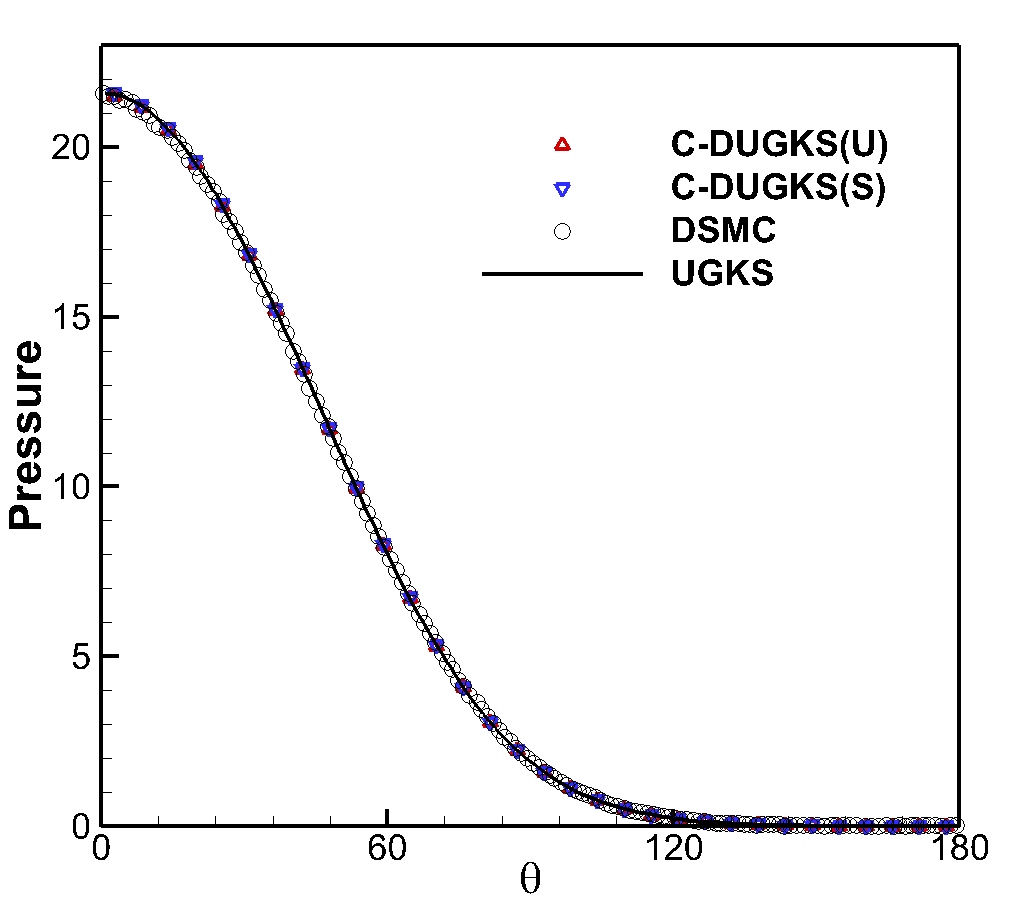}}
	\subfigure[Shear stress]{\label{fig:fig19c}\includegraphics[width=0.47\textwidth]{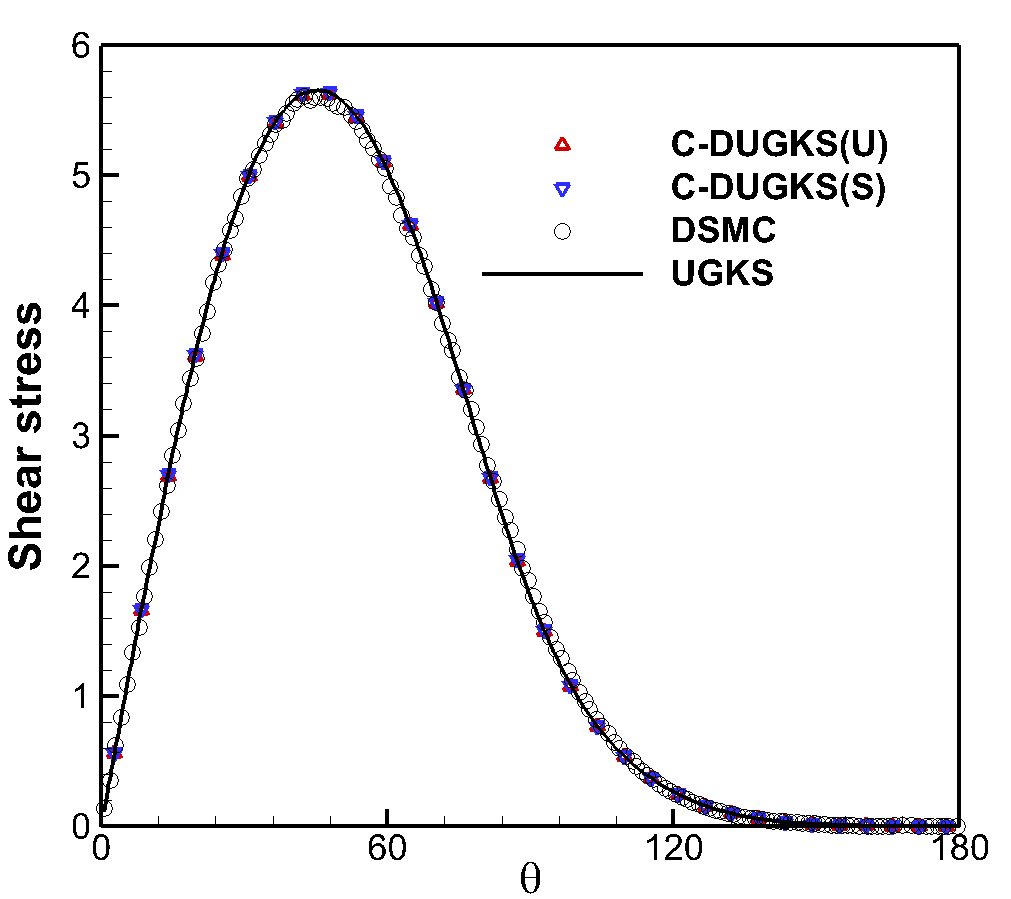}}
	\caption{\label{fig:fig19} Flow variables along the surface of the cylinder for the case of $Kn = 1.0$ (C-DUGKS(U) and C-DUGKS(S) represent C-DUGKS with unstructured and structured velocity mesh, respectively).}
\end{figure}

The flow variables along the stagnation line in front of the cylinder in Kn = 0.1 case are illustrated in Fig. \ref{fig:fig16}. The results of C-DUGKS with both structured and unstructured mesh match well the UGKS and DSMC results in Ref. \cite{Huang2013A}. In Fig. \ref{fig:fig16c}, the temperature profile predicted by both the UGKS and the C-DUGKS deviate slightly from the DSMC result in the front of the bow shock. This deviation is due to the chosen of Shakhov model equation \cite{zhu2016discrete}, and the same phenomenon can be found in the profiles of the normal shock structures \cite{liu2014investigation}. Fig. \ref{fig:fig17} are the profiles of the heat flux, pressure and shear stress on the surface of a cylinder. It is found that the results of C-DUGKS  match well with those of UGKS and DSMC. For Kn=1 case, the flow variables along the stagnation line in front of the cylinder are illustrated in Fig. \ref{fig:fig18}; the heat flux, pressure and shear stress on the surface of a cylinder are illustrated in Fig. \ref{fig:fig19}. Being the same with the Kn=0.1 case, the C-DUGKS results for Kn=1 case match well with those from the UGKS and DSMC methods. As the increasing of Kn number, the non-equilibrium effect becomes stronger, and the deviations of C-DUGKS and UGKS temperature profiles from the DSMC one becomes more obvious. Table \ref{tab:cylinder} is the convergence time of C-DUGKS with both structured and unstructured mesh. It is found that the speed-up ratios benefiting from the unstructured velocity space are 2.6 and 2.5 for Kn 0.1 and 1, respectively, which are almost the same as the ratio of discrete velocity points.

\begin{table}\label{tab:cylinder}
	\centering 
	\caption{Convergence time of C-DUGKS with both structured and unstructured velocity meshes (in seconds) for hypersonic cylinder case.}
	\begin{tabular}{p{122pt} p{60pt}<{\raggedleft} p{60pt}<{\raggedleft} p{30pt}<{\raggedleft}  }
		\hline
		\hline
		& Structured mesh (A) & Unstructured mesh (B) & ratio (A/B)\\
		\hline
		Discrete velocity points & 7921 & 3045 & 2.6\\
		\hline
		Convergence time (Kn=0.1) & 45664.0		 & 17596.4 & 2.6 \\
		\hline
		Convergence time (Kn=1) & 13352.4        & 5312.2  & 2.5 \\
		\hline
		\hline
	\end{tabular}
\end{table}

%%%%%%%%%%%%%%%%%
\section{Conclusion}\label{Conclusions}
In this paper, the coupled mass and energy distributions are introduced into the isothermal C-DUGKS, expanding its scope of simulation to non-isothermal and compressible flows in all flow regimes. Moreover, to relieve the dimensional crisis adjoint to discrete velocity space of the C-DUGKS method, a unstructured velocity space with triangular elements, which is developed for the UGKS method earlier, are introduced in this paper. The validity of the present method is proved by a series of test cases with a wide range of Kn numbers, including the Couette flow, lid-driven cavity flow, two-dimensional rarefied Rieman problem, and hypersonic flow over a circular cylinder. Especially, the validity of the present method for compressible and non-isothermal flows are verified by its precise prediction of two-dimensional rarefied Riemann problem and hypersonic cylinder flows. One important advantage of the structured uniform mesh is that high order integration methods can be used, leading to less discrete points. While, since the present unstructured velocity mesh can adjust the mesh density wherever the physical problems needs, the number discrete velocity used can be further reduced, and is significantly less than that of the structured mesh with high order integration methods. This improvement leads to a large speed-up in the two-dimensional cases of this paper. For high Mach number, this speed-up ratio will further increase, since the unstructured mesh can be made dense easily for high speed molecules concentrated in a small region of velocity space (such as hypersonic inflow).

%%%%%%%%%%%%%%%%%%%%%%%%%%%
\section*{Acknowledgements}
The authors thank Prof. Kun Xu in Hong Kong University of Science and Technology and Porf. Zhaoli Guo in Huazhong University of Science and Technology for discussions of the UGKS, the DUGKS and multi-scale flow simulations. Jianfeng Chen thanks Dr. Yong Wang in Northwestern Polytechnical University for his help in developing C-DUGKS solver on Code-Satrune. Jianfeng Chen thanks Dr. Ruifeng Yuan in Northwestern Polytechnical University for useful discussions on the unstructured velocity space.

This work has been financially supported by National Natural Science Foundation of China (No. 11472219),   111 project of China (No. B17037) and National Key Laboratory Foundation (No. G2018KC0118D).

\clearpage

%----------------------------------------------------------
% 4. References
%----------------------------------------------------------

%\bibliographystyle{elsarticle-num}
\bibliographystyle{yuan_mimplicit}
\bibliography{chen_cdugks}

\clearpage
%figures

\clearpage

\clearpage

\end{document}